\documentclass{sig-alternate_tight}
\usepackage{subfigure}
\usepackage{graphicx}
\usepackage{amsmath}
\usepackage{bm}
\usepackage{url}
\usepackage{stmaryrd}
\usepackage{balance}
\usepackage{amssymb}
\usepackage{pgfplots}
\usepackage{pifont}
\usepackage[usenames,dvipsnames]{pstricks}
\usepackage{epsfig}
\usepackage{booktabs}
\usepackage{pgffor}
\usepackage{tikz}
\usepackage{multirow}
\usepackage{amsmath}

\makeatletter
\newcommand\xleftrightarrow[2][]{%
  \ext@arrow 9999{\longleftrightarrowfill@}{#1}{#2}}
\newcommand\longleftrightarrowfill@{%
  \arrowfill@\leftarrow\relbar\rightarrow}
\makeatother

\newcommand{\mb}{\mathbf}

\newcommand{\our}{\textsc{Create}}
\newcommand{\agony}{\textsc{ASD}}
\newcommand{\oursm}{\textsc{Create-sm}}
\newcommand{\oursl}{\textsc{Create-sl}}
\newcommand{\ourl}{\textsc{Create-l}}
\newcommand{\ours}{\textsc{Create-s}}

\newcommand{\problem}{\textsc{IOC}}

\usepackage{algorithm}
\usepackage{algorithmic}

\begin{document}
\title{Organizational Chart Inference}
\numberofauthors{3}
\author{
\alignauthor
Jiawei~Zhang\thanks{This work was partially done when the first author was on a summer internship at Microsoft Research.}\\
      \affaddr{University of Illinois at Chicago}\\
      \affaddr{Chicago, IL, USA}\\
       \email{jzhan9@uic.edu}
\alignauthor
Philip~S.~Yu\\
      \affaddr{University of Illinois at Chicago, Chicago, IL, USA}\\
      \affaddr{Institute for Data Science, Tsinghua University, China}\\
       \email{psyu@cs.uic.edu}
\alignauthor
Yuanhua~Lv\\
       \affaddr{Microsoft Research}\\
       \affaddr{Redmond, WA, USA}\\
       \email{yuanhual@microsoft.com}
}
\maketitle

\begin{abstract}


Nowadays, to facilitate the communication and cooperation among employees, a new family of online social networks has been adopted in many companies, which are called the ``enterprise social networks'' (ESNs). ESNs can provide employees with various professional services to help them deal with daily work issues. Meanwhile, employees in companies are usually organized into different hierarchies according to the relative ranks of their positions. The company internal management structure can be outlined with the organizational chart visually, which is normally confidential to the public out of the privacy and security concerns. In this paper, we want to study the {\problem} (\underline{I}nference of \underline{O}rganizational \underline{C}hart) problem to identify company internal organizational chart based on the heterogeneous online ESN launched in it. {\problem} is very challenging to address as, to guarantee smooth operations, the internal organizational charts of companies need to meet certain structural requirements (about its depth and width). To solve the {\problem} problem, a novel unsupervised method {\our} (\underline{C}h\underline{A}r\underline{T} \underline{RE}cov\underline{E}r) is proposed in this paper, which consists of $3$ steps: (1) social stratification of ESN users into different social classes, (2) supervision link inference from managers to subordinates, and (3) consecutive social classes matching to prune the redundant supervision links. Extensive experiments conducted on real-world online ESN dataset demonstrate that {\our} can perform very well in addressing the {\problem} problem.

\end{abstract}


\section{Introduction}\label{sec:introduction}
\begin{figure}[t]
\centering
    \begin{minipage}[l]{0.8\columnwidth}
      \centering
      \includegraphics[width=1.0\textwidth]{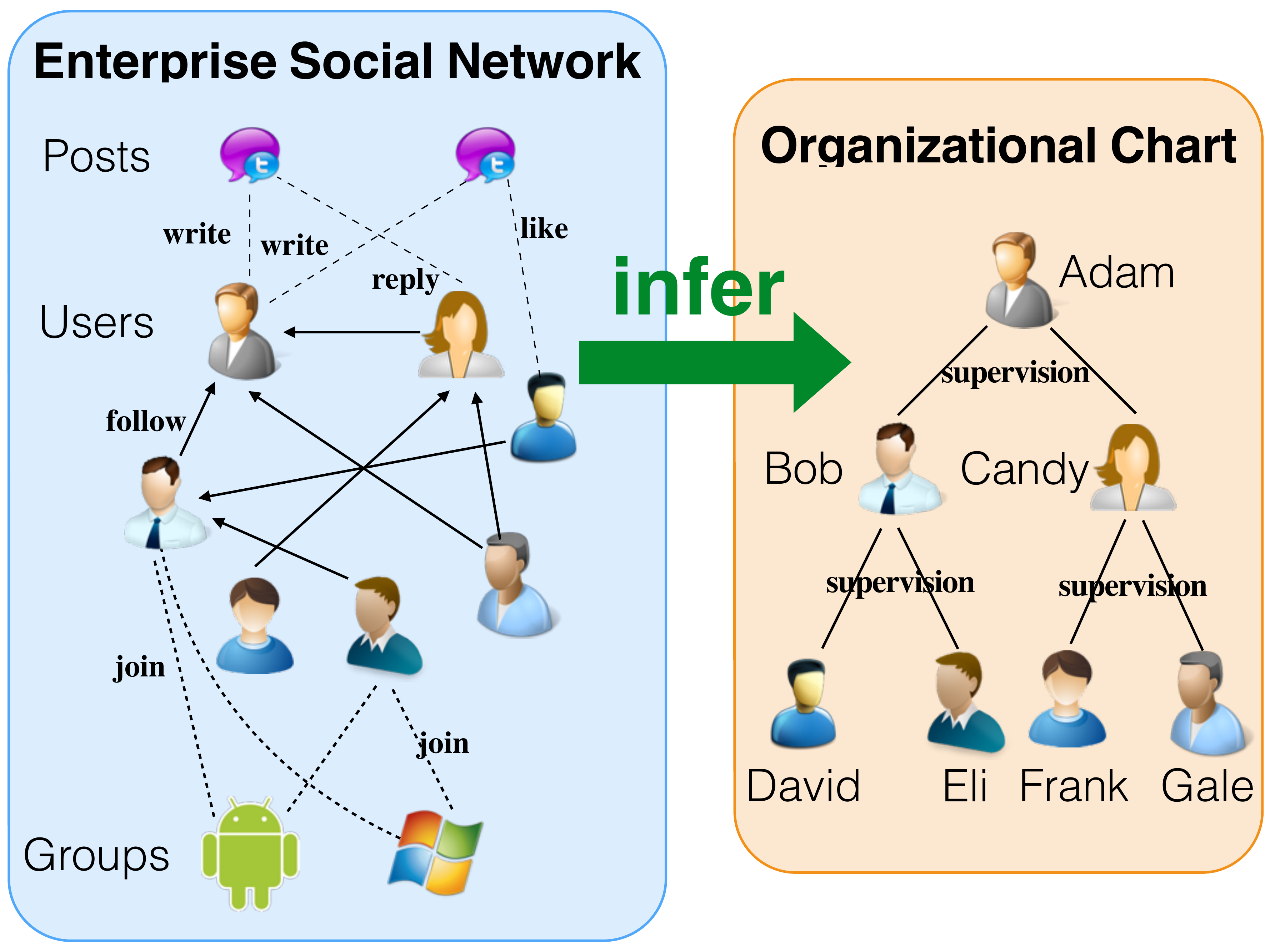}
    \end{minipage}
\caption{An example of organizational chart inference from online ESN.}\label{fig:example}
\end{figure}

In social sciences, people in social organizations (e.g., a country or a company) can be categorized into different rankings of socioeconomic tiers based on factors like wealth, income, social status, occupation, power, etc. In this paper, we will take ``company'' as an example and the internal hierarchical structure of employees in a company can be outlined with \textit{company organizational chart} formally. Most company organizational charts are usually tree-structure diagrams with CEO at the root, Executive Vice Presidents (EVPs) at the second level and so forth. Company organizational chart shows the company internal management structure as well as the relationships and relative ranks of employees with different positions/jobs, which is a common visual depiction of how a company is organized.

Nowadays, to facilitate the collaboration and communication among employees, a new type of online social networks named \textit{enterprise social networks} (ESNs) has been adopted inside the firewalls of many corporations. A representative example of online ESNs is Yammer\footnote{https://www.yammer.com/}. Over 500,000 businesses around the world are now using Yammer, including 85\% of the Fortune 500\footnote{https://about.yammer.com/why-yammer/}. Yammer provides employees with various enterprise social network services to help them deal with daily work issues and contains abundant heterogeneous information generated by employees' online social activities.


\noindent \textbf{Problem Studied}: Company internal organizational chart is usually confidential to the public for the privacy and security reasons. In this paper, we want to infer the organizational chart of a company based on the heterogeneous information in online ESNs launched in the company, and the problem is formally named as the \textit{\underline{I}nference of \underline{O}rganization \underline{C}hart} ({\problem}) problem.

To help illustrate the {\problem} problem more clearly, we also give an example in Figure~\ref{fig:example}, where the left plot is about an online ESN adopted in a company and the right plot shows the company's organizational chart. In the ESN, users can have different types of social activities, e.g., follow other users, join groups, write/reply/like posts, etc. Meanwhile, in the organizational chart, employees are connected by supervision links from managers to subordinates, who are organized into a rooted tree of depth $2$ with CEO ``Adam'' at the root. In companies, managers can usually supervise several subordinates simultaneously, while each subordinate only reports to one single manager. For instance, in Figure~\ref{fig:example}, CEO ``Adam'' manages ``Bob'' and ``Candy'' concurrently, while ``David'' only needs to report to ``Bob''.

The {\problem} problem is an interesting yet important problem. Besides inferring company organizational chart, it can also be applied in other real-world concrete applications: (1) identifying the command structures of terrorist organizations \cite{US07} based on the communication/traffic networks of their members. The command structures of terrorist organizations are usually pyramid diagrams outlining their support systems consisting of the \textit{leaders}, \textit{operational cadre}, \textit{active supporters} and \textit{passive supporters} \cite{US07}. Uncovering their internal operational structure and determining roles of members will be helpful for conducting precise strikes against their key leaders and avoid the tragic events, like 9/11 \cite{K02}. (2) inferring the social hierarchies of animals based on their observed interaction networks \cite{MB09}. Many animals (like, mammals, birds and insect species) are usually organized into dominance hierarchies. Identifying and understanding the organizational hierarchies of animals will be helpful to design and carry out effective conservation measures to protect them.

Albeit its importance, {\problem} is a novel problem and we are the first to propose to study it based on online ESNs. The {\problem} problem is totally different from existing works: (1) ``\textit{hierarchy detection in social networks}'' \cite{GSLMI11}, which only studies the division of the regular users of the social networks into different hierarchies, who are not actually involved in any organizations; (2) ``\textit{organizational intrusion}'' \cite{EFKE12}, which focuses on attacking organizations and attaining company internal information only; and (3) ``\textit{inferring offline hierarchy from social networks}'' \cite{JWPH14}, which merely infers fragments of offline hierarchical ties in \textit{homogeneous} networks, instead of reconstructing the whole organizational chart. Different from all these works, in this paper, we aim at recovering the complete organizational chart of a company (including both the hierarchical tiers of employees and the supervision links from managers to subordinates) based on the \textit{heterogeneous} information about the employees in online ESNs. 


\begin{figure}[t]
\centering
    \begin{minipage}[l]{0.8\columnwidth}
      \centering
      \includegraphics[width=1.0\textwidth]{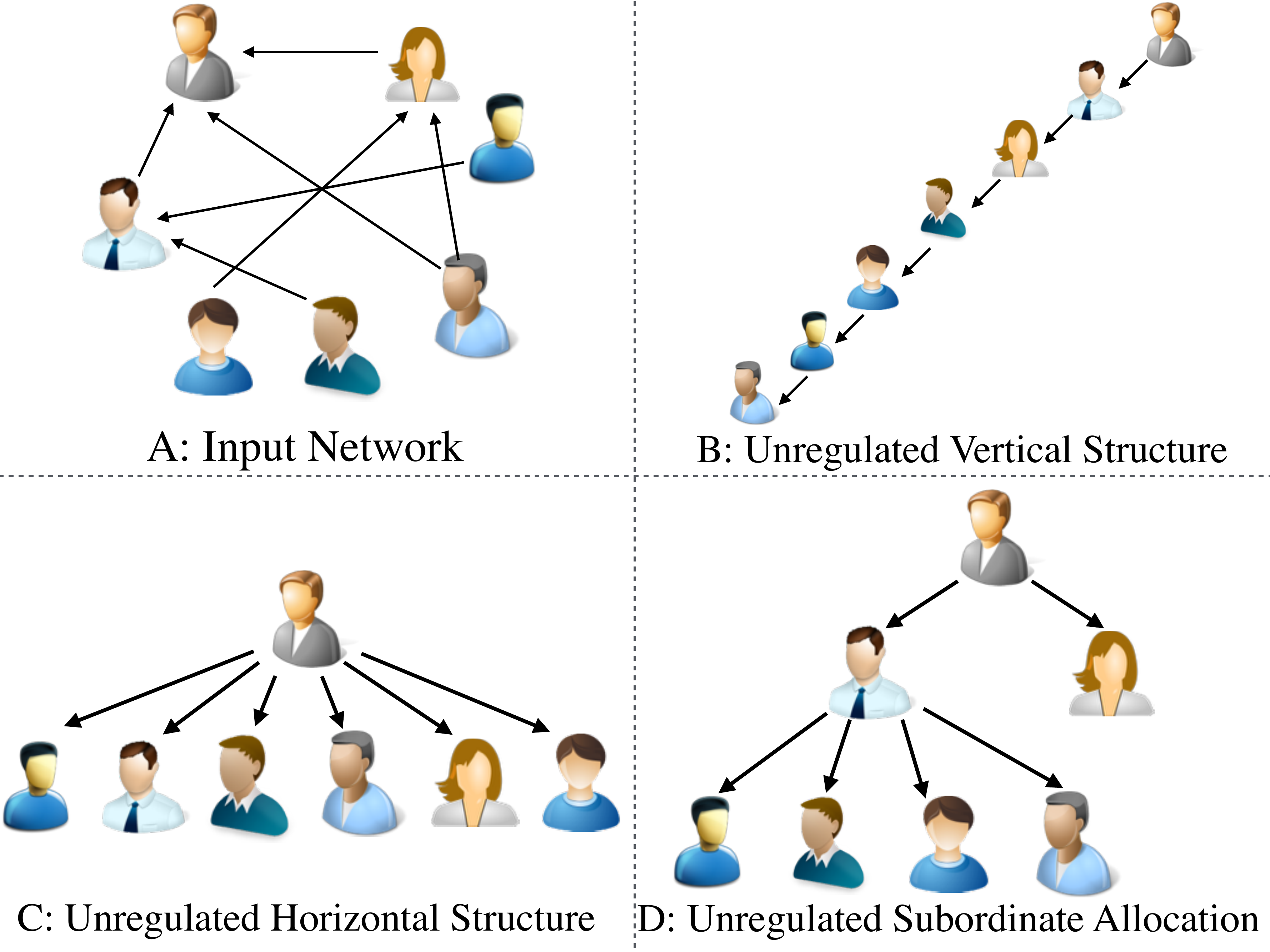}
    \end{minipage}
\caption{Examples of organizational charts.}\label{fig:balance_example}
\end{figure}

Meanwhile, to guarantee the smooth operations of companies, the inferred organizational chart needs to meet certain structural requirements \cite{org}, including both (1) macro-level depth requirement, and (2) micro-level width requirement. Two classical organizational structures adopted by companies are the vertical structure and the horizontal structure \cite{A09}. Vertical organizational structure with well-defined chains of command clearly outlines the responsibilities of each employee but will result in delays in information delivery \cite{A09}. Meanwhile, horizontal organizational structure with flat command system involves everyone in decision making but will lead to difficulties in coordinating the activities of different departments \cite{A09}. For instance, based on the input social network shown in plot A of Figure~\ref{fig:balance_example}, we give two extreme cases of the vertical and horizontal organizational charts without depth regulation in plots B and C respectively, both of which will lead to serious management problems for large companies involving tens of thousands employees. Proper regulation of the inferred organizational chart's depth (i.e., the macro-level depth requirement) is generally desired. On the other hand, most employees in companies need good supervisors to coach and instruct their daily work, but the number of subordinates each manager can supervise is limited, which can be determined by their management capacities, available time and energy. Rationally regulating the allocation of supervision workload among managers (i.e., the micro-level width requirement) can improve the management effectiveness significantly. For instance, in plot D of Figure~\ref{fig:balance_example}, we show an inferred organizational chart with depth regulation but no subordinate allocation regulation. In the plot, users in ESN are stratified into $3$ tiers (which is relatively reasonable compared to the extreme cases in plots B and C) but the employees' management workloads at tier $3$ are all assigned to one single manager, which may be beyond his/her management ability.

Despite its importance and novelty, the {\problem} problem is very hard to solve due to the following challenges:

\begin{itemize}
\vspace{-2ex}
\item \textit{regulated social stratification}: \textit{Effective social stratification} to partition users into different hierarchical arrangements (i.e., identifying the relative manager-subordinate roles of employees) while meeting the macro-level depth requirement is the prerequisite for addressing the {\problem} problem.
\vspace{-2ex}
\item \textit{supervision link inference}: Supervision link is a new type of link merely existing from managers to their subordinates. Predicting the existence of potential \textit{supervision links} with the heterogeneous information in ESNs is still an open problem.
\vspace{-2ex}
\item \textit{regulated supervision workload allocation}: To maximize the management effectiveness and efficiency, the number of subordinates each manager can supervise is limited by the management threshold $K$. In other words, supervision links in organizational chart have an inherent \textit{K-to-one} constraint.
\vspace{-2ex}


\end{itemize}

To address all the above challenges, a new unsupervised organizational chart inference framework named {\our} (\underline{C}h\underline{A}r\underline{T} \underline{RE}cov\underline{E}r) is proposed in this paper. Several new concepts (e.g., ``\textit{class transcendence social links}'', ``\textit{Matthew Effect based constraint}'' and ``\textit{chart depth regulation constraint}'') will be introduced and {\our} resolves the \textit{regulated social stratification} challenge by minimizing the existence of class transcendence social links in ESNs. {\our} tackles the \textit{supervision link inference} challenge by aggregating multiple social meta paths in the ESN between to consecutive social hierarchies. Finally, {\our} handles the \textit{regulated supervision workload allocation} challenge by applying \textit{network flow} to match consecutive social hierarchies to preserve the \textit{K-to-one} constraint on \textit{supervision links}.

The remaining parts of the paper are organized as follows. In Section~\ref{sec:formulation}, we will define some important terminologies and the {\problem} problem. Method {\our} will be introduced in Section~\ref{sec:method}. Extensive experiment results are available in Section~\ref{sec:experiment}. Finally, we describe the related works in Section~\ref{sec:relatedwork} and conclude this paper in Section~\ref{sec:conclusion}.

\section{problem formulation} \label{sec:formulation}

In this section, we will introduce the formal definitions of ``\textit{heterogeneous social network}'' and ``\textit{organizational chart}'' at first and then define the {\problem} problem with these two concepts.

\subsection{Terminology Definition}

\noindent \textbf{Definition 1} (Heterogeneous Social Networks): A \textit{heterogeneous social network} can be represented as $G = (\mathcal{V}, \mathcal{E})$, where $\mathcal{V} = \bigcup_i \mathcal{V}_i$ and $\mathcal{E} = \bigcup_i \mathcal{E}_i$ are the sets of different types of nodes and complex links among these nodes in the network respectively.

As introduced in Section~\ref{sec:introduction}, users in online ESNs (e.g., Yammer) have various types of social activities, e.g., follow other users, join groups, write/reply/like online posts, etc. As a result, Yammer can be represented as a \textit{heterogeneous social network} $G = (\mathcal{V}, \mathcal{E})$, where $\mathcal{V} = \mathcal{U} \cup \mathcal{G} \cup \mathcal{P}$  is the set of user, group and post nodes in $G$ and $\mathcal{E} = \mathcal{S} \cup \mathcal{J} \cup \mathcal{W} \cup \mathcal{R} \cup \mathcal{K}$ denotes the set of social links among users, join links between users and groups, as well as write, reply and like links between users and posts respectively.

\noindent \textbf{Definition 2} (Organizational Chart): The \textit{organization chart} of a company can be represented as a \textit{rooted tree} \cite{D97} $T = (\mathcal{N}, \mathcal{L}, root)$, where $\mathcal{N}$ is the set of \textit{employees}, $\mathcal{L}$ denotes the set of directed \textit{supervision links} from managers to subordinates in $T$ and \textit{root} represents the CEO in the company. 


\begin{figure}[t]
\centering
    \begin{minipage}[l]{0.8\columnwidth}
      \centering
      \includegraphics[width=1.0\textwidth]{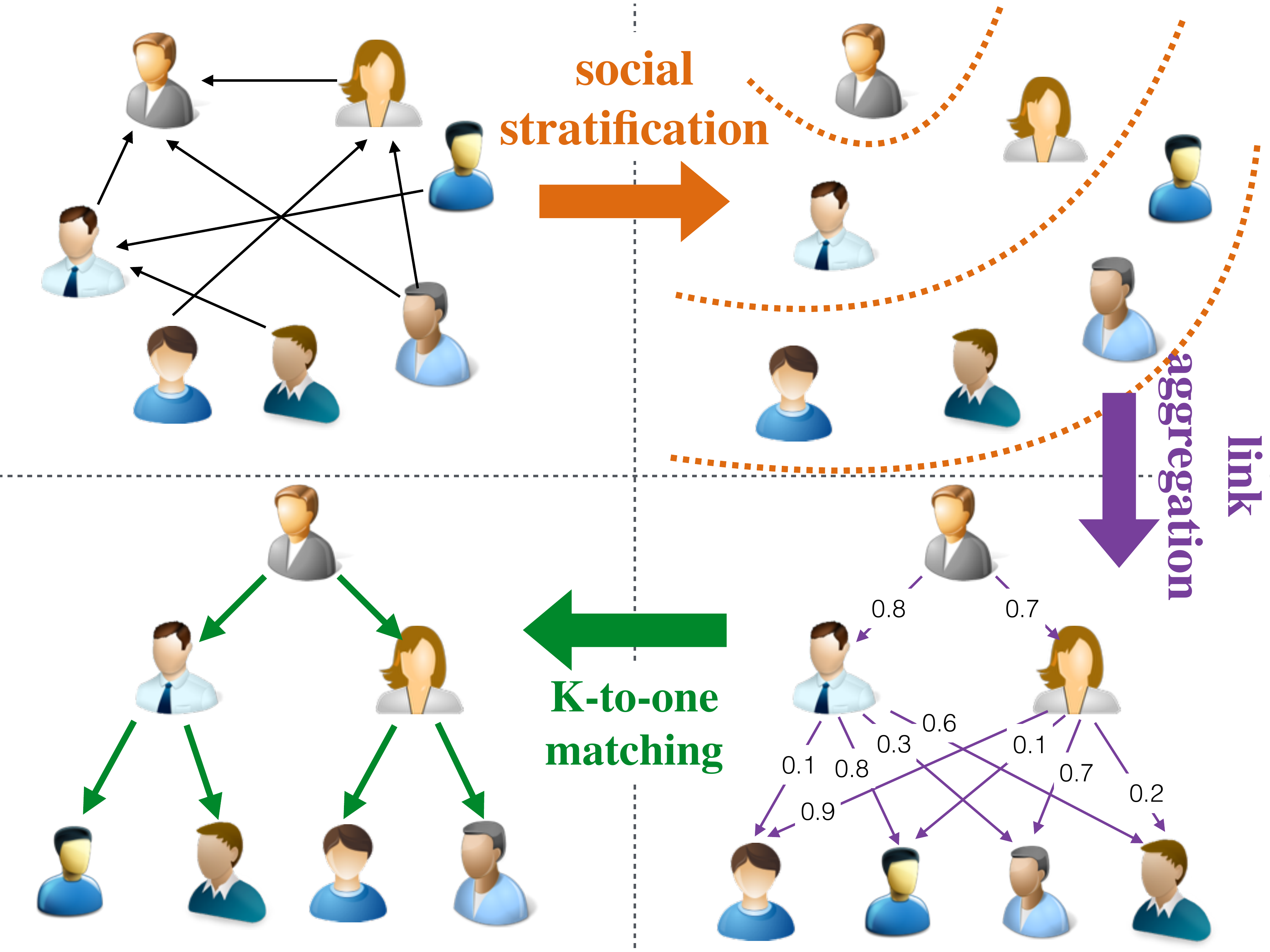}
    \end{minipage}
\caption{The framework of {\our}.}\label{fig:framework}
\end{figure}

\subsection{Problem Definition}

Based on the definitions of \textit{heterogeneous social network} and \textit{organizational chart}, we can define the {\problem} problem formally as follows:

\noindent \textbf{Definition 3} (Organizational Chart Inference ({\problem})): Given an online ESN $G = (\mathcal{V}, \mathcal{E})$ launched in a company, the {\problem} problem aims at inferring the most likely organizational chart $T = (\mathcal{N}, \mathcal{L}, root)$ of the company, where $\mathcal{N} = \mathcal{U}$ ($\mathcal{U}$ is the user set in network $G$). Furthermore, considering that the node set as well as the root node in $T$ are fixed, the {\problem} problem actually aims at inferring the $\left | \mathcal{N} -1 \right |$ most likely \textit{supervision links} $\mathcal{L}$ among employees. The inferred supervision links together with the node set $\mathcal{N}$ as well as the $root$ node can recover the original organizational chart $T = (\mathcal{N}, \mathcal{L}, root)$ of the company.

%
%


\section{Proposed Methods}\label{sec:method}

Considering that supervision links exist merely between managers and subordinates, we propose to stratify users in enterprise social networks into different \textit{social classes} to identify their relative manager-subordinate roles in Subsection~\ref{subsec:stratification}. Macro-level depth requirement of the inferred chart is achieved with the \textit{depth regulation constraint} in the social stratification objective function. Potential supervision links can be inferred between employees in consecutive classes by aggregating social meta paths among employees in the ESN in Subsection~\ref{subsec:metapath}. To preserve the \textit{K-to-one} constraint on supervision links (i.e., the micro-level width requirement), redundant non-existing supervision links will be pruned in Subsection~\ref{subsec:networkflow}. Generally, as shown in Figure~\ref{fig:framework}, framework {\our} has three steps: (1) regulated social stratification, (2) supervision link inference, and (3) regulated social class matching, which will all be introduced in this section.

\subsection{Regulated Social Stratification}\label{subsec:stratification}

Supervision links merely exist between managers and subordinates. Division of users into hierarchies to identify their relative manager-subordinate roles can shrink the supervision link inference space greatly. The process of hierarchizing users in online ESN is called \textit{social stratification} formally.


\noindent \textbf{Definition 4} (Social Stratification): Traditional \textit{social stratification} concept used in social science denotes the ranking and partition of people into different hierarchies based on various factors, e.g., power, wealth, knowledge and importance \cite{W13}. In this paper, we define \textit{social stratification} as the partition process of users in online ESNs into different hierarchies according to their management relationships, where managers are at upper levels, while subordinates are at lower levels.


The relative stratified levels of users in online ESN are defined as their social classes.

\noindent \textbf{Definition 5} (Social Class): \textit{Social class} is a term used by social stratification models in social science and the most common ones are the upper, middle, and lower classes \cite{G01}. In this paper, we define \textit{social class} of users in online ESNs as their management level in the company, where CEO belongs to social class 1, EVPs belong to class 2, and so forth.

In social stratification, users in online ESNs will be mapped to their \textit{social classes} according to mapping: $c: \mathcal{U} \to \mathbb{Z}^+$. For each user $u \in \mathcal{U}$, his social class $c(u)$ is defined recursively as follows:
$$
c(u) = 
\begin{cases} 
1,  & \mbox{if }u\mbox{ is the CEO;} \\
c\left (m(u) \right) + 1, & \mbox{otherwise.}
\end{cases}
$$
where $m(u)$ represents the direct manager of $u$.

In social science, the working class are eager to get acquainted with and join the upper echelons of their class by either accumulating wealth \cite{life}, imitating their dressing styles \cite{fasion}, and mimicking their dialect and accents \cite{K95}. Meanwhile, the upper class are very cohesive and they tend to be friends who share similar background \cite{H12}. So is the case for the social links in \textit{enterprise social networks}. By analyzing the Yammer network data, we observe that the probability for users to follow upper-level managers is $31.9\%$ on average, while that of following subordinates is merely $11.2\%$. As a result, in online ESNs, subordinates tend to follow their managers, while people in management are reluctant to initiate the friendship with their subordinates \cite{boss}. Based on such an observation, we introduce the concept of \textit{class transcendence} social links and propose to stratify users by minimizing the existence of such links in ESNs.

\noindent \textbf{Definition 6} (Class Transcendence Social Link): Link $(u, v)$ (i.e., $u$ follows $v$) is defined as a \textit{class transcendence social link} in online ESN $G$ iff $(u, v) \in \mathcal{S}$ and $c(u) < c(v)$ (where smaller social class denotes upper management level in the organizational chart).


In social stratification, each introduced class transcendence social link in the result will lead to a \textit{class transcendence penalty}. Let $c(\mathcal{U}) = \{c(u_1), c(u_2), \cdots, c(u_{\left | \mathcal{U} \right |})\}$ be the \textit{social stratification} result of all users in the ESN. For any directed social link $(u, v) \in \mathcal{S}$ in the ESN, the \textit{class transcendence penalty} introduced by it can be represented as 
$$
p\left(c(u), c(v)\right) = \begin{cases} 
0,  & \mbox{if }c(u) > c(v) \\
c(v) - c(u) + 1, &\mbox{otherwise.}
\end{cases}
$$

The \textit{class transcendence penalty} introduced by all social links (i.e., $\mathcal{S}$) in the ESN can be represented as
\begin{align*}
p\left(c(\mathcal{U})\right) &= \sum_{(u, v) \in \mathcal{S}} p\left(c(u), c(v)\right)\\
&= \sum_{(u, v) \in \mathcal{S}} \max \{c(v) - c(u) + 1, 0\}.
\end{align*}

%

``The rich get richer" (i.e., the Matthew Effect \cite{M88}) is a common phenomenon in social science literally referring to issues of fame or status as well as cumulative advantage of economic capital. By analyzing the Yammer network data, we have similar observations: ``people at higher management level can accumulate more followers easily''. Such an observation provides important hints for inferring users' relative management levels according to their in degrees in ESN (i.e., the number of followers).

\noindent \textbf{Definition 7} (Matthew Effect based Constraint): For any two given users $u$ and $v$ in the network, let $\Gamma(u)$ and $\Gamma(v)$ be the follower sets of $u$ and $v$ in the network respectively. The \textit{matthew effect based constraint} on users $u$ and $v$ can be represented as $c(u) \le c(v)$ if $|\Gamma(u)| \ge |\Gamma(v)|$.

Furthermore, to maximize the operation efficiency of companies, the inferred organizational chart needs to meet the \textit{macro-level depth requirement}, which can be achieved with the following chart depth regulation constraint.


\noindent \textbf{Definition 8} (Chart Depth Regulation Constraint): The \textit{chart depth regulation constraint} avoids obtaining organizational chart with too short command chains (e.g., the extreme horizontal structure) and can be represented as
$$\sum_{u \in \mathcal{U}} c(u) \ge \alpha \cdot \left | \mathcal{U} \right |,$$
where parameter $\alpha$ is used to regulate the depth of the chart, whose sensitivity analysis will be given in Section~\ref{sec:experiment}. Furthermore, term $\sum_{u \in \mathcal{U}} c(u)$ is also added to the minimization objective function to avoid obtaining charts with too long command chains (i.e., the extreme vertical structure).

Based on all the above remarks, the \textit{optimal} regulated social stratification $c^*(\mathcal{U})$ of users in ESN can be obtained by solving the following objective function:
\begin{align*}
c^*(\mathcal{U}) &= \arg \min_{\{c(u_1), c(u_2), \cdots, c(u_{\left | \mathcal{U} \right |})\}} \sum_{(u, v) \in \mathcal{S}} p\left(c(u), c(v)\right) + \sum_{u \in \mathcal{U}} c(u),\\
s.t.,\ \  &p(c(u), c(v)) \ge c(v) - c(u) + 1, \forall (u, v) \in \mathcal{S},\\
&p(c(u), c(v)) \ge 0, \forall (u, v) \in \mathcal{S},\\
&c(u) \le c(v), \forall u,v \in \mathcal{U}, \mbox{ if } |\Gamma(u)| \ge |\Gamma(v)|,\\
&\sum_{u \in \mathcal{U}} c(u) \ge \alpha \cdot \left | \mathcal{U} \right |,\\
&c(u) = 1, \mbox{ if } u \mbox{ is the CEO},\\
&c(u) > 1, c(u) \in \mathbb{Z}^+, \forall u \in \mathcal{U}\setminus\{\mbox{CEO}\},\\
&p\left(c(u), c(v)\right) \in \mathbb{Z}, \forall (u, v) \in \mathcal{S}.
\end{align*}

The integer programming objective function can be solved with open source toolkits, e.g., GLPK\footnote{https://www.gnu.org/software/glpk/}, PuLP\footnote{https://code.google.com/p/pulp-or/}, etc., very easily and the obtained results of variables $c(u_1), c(u_2), \cdots, c(u_{\left | \mathcal{U} \right |})$ represent the inferred social classes of users in online ESN.


%

\subsection{Supervision Link Inference with Social Meta Paths Aggregation}\label{subsec:metapath}

It is a challenge to estimate the supervision relations between the ESN members in consecutive social classes. Here we use the meta paths concept introduced in \cite{SBGAH11, SHYYW11, ZYZ14} to identify and evaluate different types relationship in ESN.

\noindent \textbf{Social Meta Paths in Enterprise Social Networks}

\begin{itemize}
\vspace{-2ex}
\item \textit{Follow}: User $\xrightarrow{follow}$ User, whose notation is ``$U \to U$'' or $\Phi_1(U, U)$.
\vspace{-2ex}
\item \textit{Follower of Follower}: User $\xrightarrow{follow}$ User $\xrightarrow{follow}$ User, whose notation is ``$U \to U \to U$'' or $\Phi_2(U, U)$.
\vspace{-2ex}
\item \textit{Common Followee}: User $\xrightarrow{follow}$ User $\xrightarrow{follow^{-1}}$ User, whose notation is ``$U \to U \gets U$'' or $\Phi_3(U, U)$.
\vspace{-2ex}
\item \textit{Common Follower}: User $\xrightarrow{follow^{-1}}$ User $\xrightarrow{follow}$ User, whose notation is ``$U \gets U \to U$'' or $\Phi_4(U, U)$.
\vspace{-2ex}
\item \textit{Common Group Membership}: User $\xrightarrow{join}$ Group $\xrightarrow{join^{-1}}$ User, whose notation is ``$U \to G \gets U$'' or $\Phi_5(U, U)$.
\vspace{-2ex}
\item \textit{Reply Post}: User $\xrightarrow{write}$ Post $\xrightarrow{reply}$  Post $\xrightarrow{write^{-1}}$ User, whose notation is ``$U \to P\to P \gets U$'' or $\Phi_6(U, U)$.
\vspace{-2ex}
\item \textit{Like Post}: User $\xrightarrow{write}$ Post $\xrightarrow{like^{-1}}$ User, whose notation is ``$U \to P\to P \gets U$'' or $\Phi_7(U, U)$.
\vspace{-2ex}
\end{itemize}

An existing user intimacy \cite{ZY15} measure, Path-Sim, based on meta paths was introduced in \cite{SHYYW11}, which can calculate the propagation probability between users via meta paths in \textit{undirected homogeneous networks}. To deal with \textit{directed heterogeneous networks}, we extend it and introduce a new intimacy measure, DP-intimacy (Directed Path-Intimacy), based on social meta path $\Phi_i(U, U), i \in \{1, 2, \cdots, 7\}$:
$$\mbox{DP-intimacy}_i(u, v) = \frac{ \left| \mathcal{PATH}_i(u \rightsquigarrow v) \right| + \left| \mathcal{PATH}_i(v \rightsquigarrow u) \right| } {\left| \mathcal{PATH}_i(u \rightsquigarrow \cdot) \right|  + \left| \mathcal{PATH}_i(v \rightsquigarrow \cdot) \right|},$$
where $\mathcal{PATH}_i(u \rightsquigarrow v)$ denotes the instance set of meta path $\Phi_i(U, U)$ going from $u$ to $v$ in the ESN.


Different social meta paths capture the intimacy between users in different aspects and overall intimacy between users can be obtained by aggregating information from all these social meta paths. Let $\mbox{DP-intimacy}_1(u,v), \mbox{DP-intimacy}_2(u,v), \\ \cdots, \mbox{DP-intimacy}_7(u,v)$ be the intimacy scores between users $u$ and $v$ calculated based on social meta paths $\Phi_1(U, U),\\ \Phi_2(U, U), \cdots, \Phi_7(U, U)$ respectively. Without loss of generality, we choose logistic function as the intimacy aggregation function \cite{GSHB14}, the overall intimacy between users $u$ and $v$ can be represented as
$$intimacy(u, v)=\frac{e^{\sum\nolimits_{(i)}\omega_i \mbox{DP-intimacy}_i(u, v)}} {1+ e^{\sum\nolimits_{(i)}\omega_i \mbox{DP-intimacy}_i(u, v)}} \in [0, 1],$$
where the value of $\omega_i$ denotes the weight of social meta path $\Phi_i$ and $\sum_i \omega_i = 1$.
%
%
%
%
%

\begin{figure}[t]
\centering
    \begin{minipage}[l]{0.8\columnwidth}
      \centering
      \includegraphics[width=1.0\textwidth]{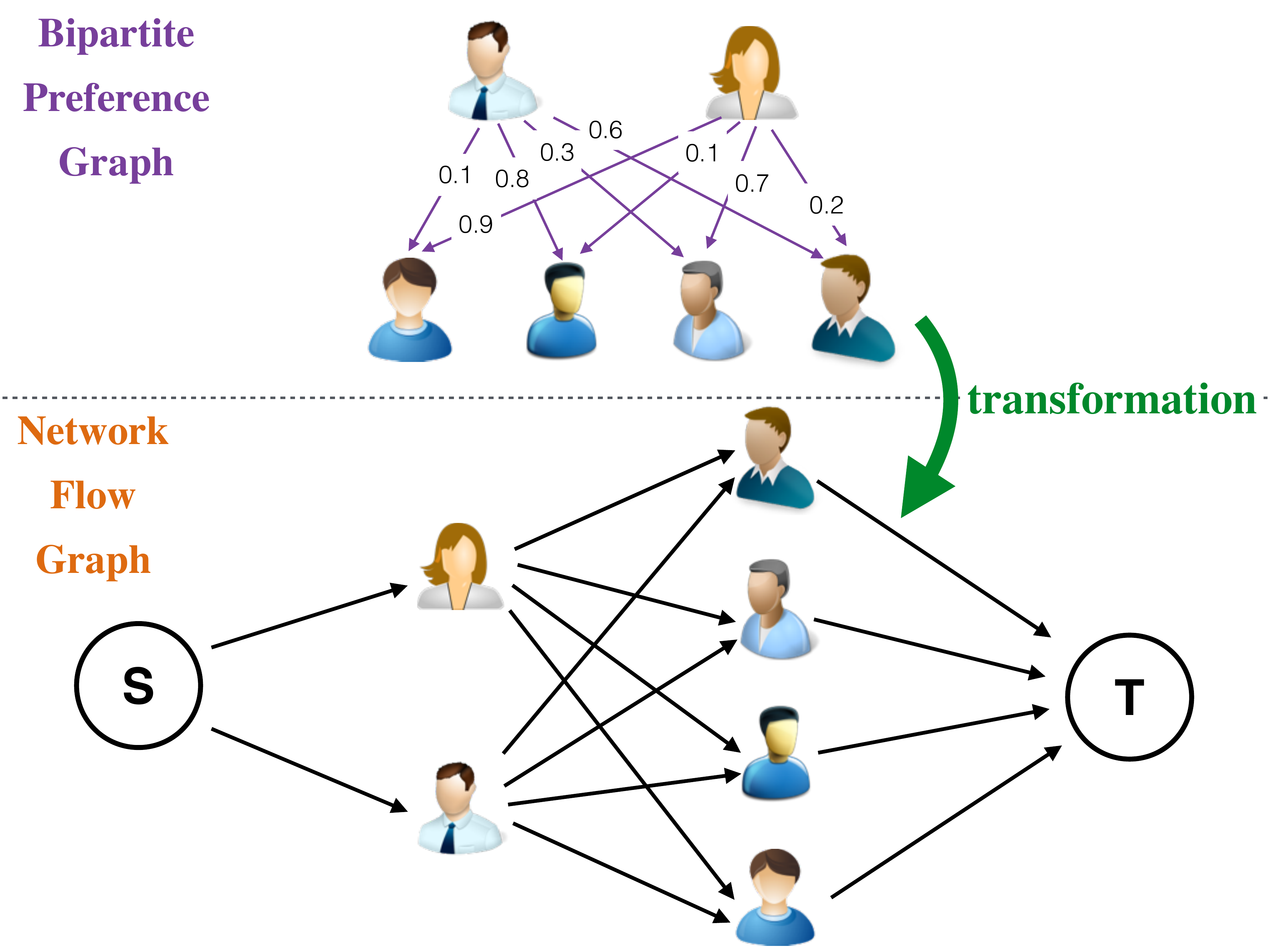}
    \end{minipage}
\caption{An example of K-to-one matching.}\label{fig:network_flow}
\end{figure}

\subsection{Regulated Social Class Matching}\label{subsec:networkflow}

Meta path aggregation based supervision link inference method proposed in previous step calculate the intimacy scores of all the potential links between pairs of social classes. However, to regulate the supervision workload allocation, the number of subordinates each manager can supervise is limited by the management threshold $K$. In this section, we will prune the redundant non-existing supervision links with network-flow based regulated social class matching to preserve the \textit{K-to-one} constraint on \textit{supervision links}.



\subsubsection{Bipartite Preference Graph}

Based on the \textit{social stratification} results $c(\mathcal{U}) = \{c(u_1), c(u_2),\\ \cdots, c(u_{\left | \mathcal{U} \right |})\}$, we can stratify all the users into \textit{social classes} $[1, \max(c(\mathcal{U}))]$ and users in class $i \in [1, \max(c(\mathcal{U}))]$ can be represented as set $\Psi(i) \subset \mathcal{U}$. By aggregating information in various \textit{social meta paths}, we can calculate the intimacy scores of all potential \textit{supervision links} between consecutive \textit{social classes}, which exist between $\Psi(i)$ and $\Psi(i+1)$ can be represented as set $\Lambda(i, i + 1) = \Psi(i) \times \Psi(i+1)$. Links in $\Lambda(i, i + 1)$ are associated with certain weights (i.e., the calculated intimacy scores), which can be obtained with the mapping $\Pi(i, i + 1): \Lambda(i, i + 1) \to \mathbb{R}$.


Users in social classes $i$ and $i+1$ (i.e., $\Psi(i)$ and $\Psi(i+1)$) together with all the potential supervision links between them (i.e., $\Lambda(i, i + 1)$) and their intimacy scores (i.e., $\Pi(i, i + 1)$) can form a weighted bipartite preference graph.


\noindent \textbf{Definition 9} (Weighted Bipartite Preference Graph): The \textit{weighted bipartite preference graph} between users in $\Psi(i)$ and $\Psi(i+1)$ can be represented as $B = (\Psi(i) \cup \Psi(i+1), \Lambda(i, i + 1), \Pi(i, i + 1))$.


An example of \textit{weighted bipartite preference graph} is shown in the upper plot of Figure~\ref{fig:network_flow}. In the example, all the potential supervision links between the upper-level and lower-level individuals are represented as the directed purple lines between them, whose weights are the numbers marked on the lines. Each employee in the figure is associated with multiple potential \textit{supervision links} and the redundant ones can be pruned with the \textit{network flow} method introduced in the next subsection.

\subsubsection{Minimum Cost Network Flow based Social Class Matching}

Based on the \textit{bipartite preference graph} $B$, we propose to construct the following \textit{network flow graph} first.

\noindent \textbf{Definition 10} (Network Flow Graph): Based on bipartite preference graph $B = (\Psi(i) \cup \Psi(i+1), \Lambda(i, i + 1), \Pi(i, i + 1))$, the \textit{network flow graph} can be represented as $H = (\mathcal{N}_H, \mathcal{L}_H, \mathcal{W}_H)$. Node set $\mathcal{N}_H$ includes all nodes in $B$ and two dummy nodes: source node $s$ and sink node $t$ (i.e., $\mathcal{N}_H = \Psi(i) \cup \Psi(i+1) \cup \{s, t\}$). Besides all the links in $B$, we further add directed links from $s$ to all nodes in $\Psi(i)$, as well as those from all nodes in $\Psi(i+1)$ to $t$ (i.e., $\mathcal{L}_H = \Lambda(i, i + 1) \cup \left(\{s\} \times \Psi(i)\right) \cup \left(\Psi(i+1) \times \{t\} \right)$). Only the links in $\Lambda(i, i + 1)$ are associated with weights, which can be obtained with mapping $\mathcal{W}_H = \Pi(i, i + 1)$.

For instance, based on the \textit{bipartite preference graph} in the upper plot of Figure~\ref{fig:network_flow}, we can construct its corresponding \textit{network flow graph} (i.e., the lower plot). All the links in the network flow graph are directed denoting the flow direction.

\noindent \textbf{Bound Constraint of Network Flow}

For each link $(u, v) \in \mathcal{L}_H$, we allow a certain amount of flow going through within range $[lb_{u, v}, up_{u, v}]$, where $lb_{u, v}$ and $up_{u, v}$ represent the \textit{lower bound} and \textit{upper bound} associated with link $(u, v)$ respectively and
$$lb_{u, v} \le x_{u, v} \le ub_{u, v},$$
where $x_{u, v}$ is the flow amount going through link $(u, v)$.

More specifically, for links from $s$ to the upper level individuals, i.e., $\{s\} \times \Psi(i)$, we set its \textit{lower bound} and \textit{upper bound} to be $lb_{s, u} = 0$ and $ub_{s, u} = K$ respectively and we can get
$$0 \le x_{s, u} \le K, \forall u \in \Psi(i),$$
where $K$ is the management threshold, whose sensitivity analysis is available in Section~\ref{sec:experiment}. It is actually the constrain to preserve micro-level width requirement.

For link $(v, t) \in \mathcal{L}_H$, we set its \textit{lower bound} and \textit{upper bound} to be $lb_{v, t} = 1$ and $ub_{v, t} = 1$, i.e.,
$$x_{v, t} = 1, \forall v \in \Psi(i+1),$$
which means exact amount $1$ flow goes through link $(v, t)$ (i.e., each subordinate needs to have exactly one manager).

Links $(u, v) \in \Lambda(i, i + 1) \subset \mathcal{L}_H$ have lower and upper bounds $lb_{u,v} = 0$, $ub_{u,v} = 1$ and the flow amount needs to be an integer, i.e., 
$$x_{u, v} \in \{0, 1\},$$
denoting whether supervision links in $\Lambda(i, i + 1)$ are selected or not in the matching result.

\noindent \textbf{Mass-Balance Constraint of Network Flow}

In network flow model, for each node, e.g., $u$, the amount of flow going into $u$ should be equal to that going out from $u$, i.e., 
$$\sum_{w \in \mathcal{N}_H, (w, u) \in \mathcal{L}_H} x_{w,u} = \sum_{v \in \mathcal{N}_H, (u, v) \in \mathcal{L}_H} x_{u, v}.$$

\noindent \textbf{Minimum Cost Network Flow}

All links going from $\Psi(i)$ to $\Psi(i+1)$ are associated with corresponding \textit{flow costs}, which are negatively correlated to their intimacy scores. For instance, in this paper, for link $(u, v) \in \mathcal{L}_B$ with weight $intimacy(u, v)$, we can represent their \textit{flow cost} as $1 - intimacy(u, v)$. The optimal \textit{network flow} with the \textit{minimum cost} (i.e., the maximum intimacy) can be obtained by addressing the following \textit{integer programming} problem:

\begin{align*}
\min &\sum_{(u, v) \in \Lambda(i, i + 1)} x_{u, v} (1 - intimacy(u, v))\\
s.t. \ \ &0 \le x_{s, u} \le K, \mbox{for }\forall u \in \Psi(i),\\
&x_{v, t} = 1, \mbox{for }\forall v \in \Psi(i+1),\\
&x_{u, v} \in \{0, 1\}, \mbox{for }\forall u \in \Psi(i), \forall v \in \Psi(i+1),\\
&\sum_{w \in \mathcal{N}_H, (w, u) \in \mathcal{L}_H} x_{w,u} = \sum_{v \in \mathcal{N}_H, (u, v) \in \mathcal{L}_H} x_{u, v}, \forall u \in \mathcal{N}_H.
\end{align*}

Similarly, the above integer programming problem can be addressed with open source toolkits and how to solve the equation will not be introduced here. Variables obtained by solving the above equation can lead to the minimum cost but can also meet the constraints as well. These obtained variables denote the existence scores of the corresponding supervision links, where the selected links (i.e., those corresponding variable $x=1$) will be assigned with label $+1$ while the rest are assigned with label $-1$.


\section{Experiments}\label{sec:experiment}

To examine the performance of {\our} in addressing the {\problem} problem, in this part, extensive experiments will be conducted on real-world \textit{enterprise social network}: Yammer. 

\subsection{Dataset Description}

We crawl all the Microsoft employees' information from Yammer and obtain the complete organizational chart involving all these employees in Microsoft during June, 2014. The social network data covers all the user-generated content (such as posts, replies, topics, etc.) and social graphs (such as user-user following links, user-group memberships, user-topic following links, etc.) by then that are set to be public. In summary, it includes more than $100k$ Microsoft employees, and millions of user-generated posts published and the social links.\footnote{We are not able to reveal the actual numbers here and throughout the paper for commercial reasons.}


%

All the users in yammer are registered with the official employment ID in Microsoft, via which we can identify them in the organizational chart correspondingly. From Microsoft, the complete organization structure of all employees is obtained. As introduced before, the structure of the organizational chart is a rooted tree with the CEO at the top.


\subsection{Experiment Settings}

The {\our} framework proposed in this paper is an unsupervised model, and the \textit{organizational chart} is used for evaluation only in the experiments. To ensure the employee node set of organizational chart to be identical to that of ESN, a fully aligned \cite{KZY13} Yammer network and organizational chart are sampled from the dataset. Initially, with the directed follow links among users in Yammer, we achieve the regulated social stratification of users by minimizing the number \textit{class transcendence social links}. All the potential supervision links between pairwise consecutive social classes in the social stratification result are inferred by aggregating information from various social meta paths in Yammer, whose existence likelihood is denoted as the intimacy score. For simplicity, the weights of different social meta paths in logistic function are assigned with identical values, i.e., $\mb{\omega} = [\frac{1}{7}, \cdots, \frac{1}{7}]$. A subset of these inferred supervision links will be selected via the regulated social class matching based on the network-flow model to preserve the \textit{K-to-one} constraint on supervision links. 

Meanwhile, to demonstrate the effectiveness of {\our}, we compare {\our} with many baseline methods, including both state-of-art and traditional methods in social stratification and organizational chart inference.

\noindent \textbf{Social Stratification Methods}:

\begin{itemize}
\vspace{-2ex}
\item \textit{Regulated Social Stratification}: Regulated social stratification is the first step of {\our} proposed in this paper, which is also named as {\our} for simplicity. {\our} exploits the concept of class transcendence social links and Matthew Effect based constraint to stratify users in ESNs into different social classes. In addition, to regulate the depth of inferred social classes about employees, {\our} further adds a \textit{chart depth regulation constraint} into the objective function. 
\vspace{-2ex}
\item \textit{Agony based Social Division}: {\agony} is a state-of-art social division method proposed in \cite{GSLMI11}, which detects the social hierarchies of regular users in general online social networks. {\agony} is not designed for organizational chart inference and doesn't consider the matthew effect based constraint nor the chart depth regulation constraint.
\vspace{-2ex}
\end{itemize}

\noindent \textbf{Organizational Chart Inference Methods}:

\begin{itemize}
\vspace{-2ex}

\item \textit{Social Stratification + Link Prediction + Matching} ({\our}): {\our} is the framework proposed in this paper and it has three steps: (1) regulated social stratification, (2) link inference and (3) regulated social class matching.
\vspace{-2ex}


\item \textit{Social Stratification + Link Prediction} ({\oursl}): {\oursl} contains two steps: (1) \textit{social stratification}, and (2) \textit{link prediction} based on accumulated social meta paths. {\oursl} has no \textit{matching step} to keep the micro-level width requirement and the outputs cannot meet the \textit{K-to-one} constraint.
\vspace{-2ex}
\item \textit{Social Stratification + Matching} ({\oursm}): {\oursm} contains two steps: (1) \textit{social stratification}, and (2) social class matching. {\oursm} has no supervision link prediction step and social links from upper-level social class to the lower-level are regarded as the potential supervision links candidates.
\vspace{-2ex}
\item \textit{Social Stratification} ({\ours}): {\ours} is identical to {\oursl} except that it has no matching step and outputs the all the \textit{social links} between sequential hierarchies as the supervision links.
\vspace{-2ex}

\item \textit{Traditional Unsupervised Link Prediction Methods}: No existing supervised link prediction models can be applied as no labeled supervision link exist. For completeness, we further compare \textit{our} with traditional unsupervised baseline methods which include \textit{Common Neighbor (CN)} \cite{HZ11}, \textit{Jaccard's Coefficient (JC)} \cite{HZ11} and \textit{Adamic Adar (AA)} \cite{AA01} between consecutive stratified social classes.
\vspace{-2ex}
\end{itemize}

\noindent \textbf{Social Stratification Evaluation Metrics}

In the \textit{social stratification} step, the outputs are the inferred \textit{social classes} of all the employees. By comparing them with individuals' real-world \textit{social classes} (i.e., the ground truth), we can calculate the \textit{mean absolute error} \cite{WBSS04}, \textit{mean squared error} \cite{WBSS04} and \textit{coefficient of determination} (i.e., $R^2$) \cite{NBZ06} of the results. In addition, the ratio of correctly stratified users (i.e., accuracy) can also be used to measure the performance. So, the metrics used to evaluate the performance of different social stratification methods include \textit{mean absolute error} (MAE) \cite{HK06}, \textit{mean squared error} (MSE) \cite{RPD88}, $R^2$ \cite{RPD88} and \textit{accuracy}.

\noindent \textbf{Organizational Chart Inference Evaluation Metrics}

Methods {\oursl}, {\ourl} and the traditional unsupervised link prediction can only output the confidence scores of all potential supervision links without labels, whose performance will be evaluated by metrics $AUC$ and Precision@100. Meanwhile, {\our} and {\oursm} can output both labels and scores of potential supervision links and, besides $AUC$ and Precision@100, we will also evaluate their performance with Precision, Recall and F1-score.

\subsection{Social Stratification Results}

\begin{figure}[t]
\centering
    \begin{minipage}[l]{0.65\columnwidth}
      \centering
      \includegraphics[width=1.0\textwidth]{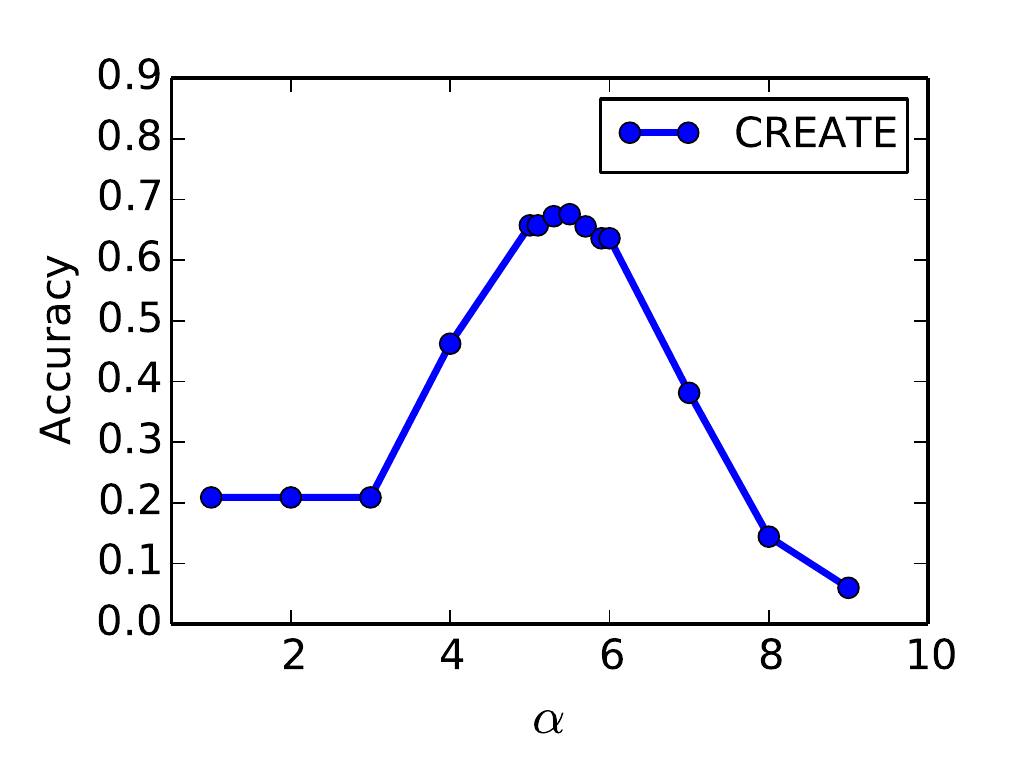}
    \end{minipage}
\caption{Sensitivity analysis of parameter $\alpha$.}\label{fig:alpha}
\end{figure}

In social stratification, parameter $\alpha$ is applied to maintain the macro-level depth requirement, which can control the depth of the organizational chart. Before comparing the performance of {\our} with {\agony}, we will analyze the sensitivity of parameter $\alpha$ at first. We select $\alpha$ with values in $\{1, 2, 3, 4, 5, 5.1, 5.3, 5.5, 5.7, 5.9, 6, 7, 8, 9\}$ and obtain the accuracy scores achieved by {\our} as shown in Figure~\ref{fig:alpha}.

When parameter $\alpha$ is very small (e.g., from $1$ to $3$), we observe that it has no effects on the performance of {\our}. The possible reason can be that \textit{Matthew Effect based constraint} can already effectively outline the relative hierarchical relationships among users in online ESNs, the average social class of users obtained based on which is already greater than $3$. When $\alpha$ becomes larger (from $4$ to $6$), the \textit{structure regulation constraint} starts to matter more and the social stratification accuracy goes up steadily and achieve the highest value at $5.5$, i.e., the default value of $\alpha$ in later experiments. {\our} performs better as $\alpha$ increases shows that the \textit{structure regulation constraint} can stretch the organizational structure and stratify users in their correct social classes. However, as $\alpha$ further increases (i.e., from $6$ to $9$), the accuracy achieved by {\our} decreases dramatically. The reason can be that larger $\alpha$ stretches the organizational structure too much and put lots users into the wrong social classes. For example, it is nearly impossible for users to achieve $9$ as the average social class, which is actually the largest social class in the sampled fully aligned organizational chart.

\begin{figure*}[t]
\centering
\subfigure[Precision]{\label{eg_fig_comp_level_1}
    \begin{minipage}[l]{1.0\columnwidth}
      \centering
      \includegraphics[width=1.0\textwidth]{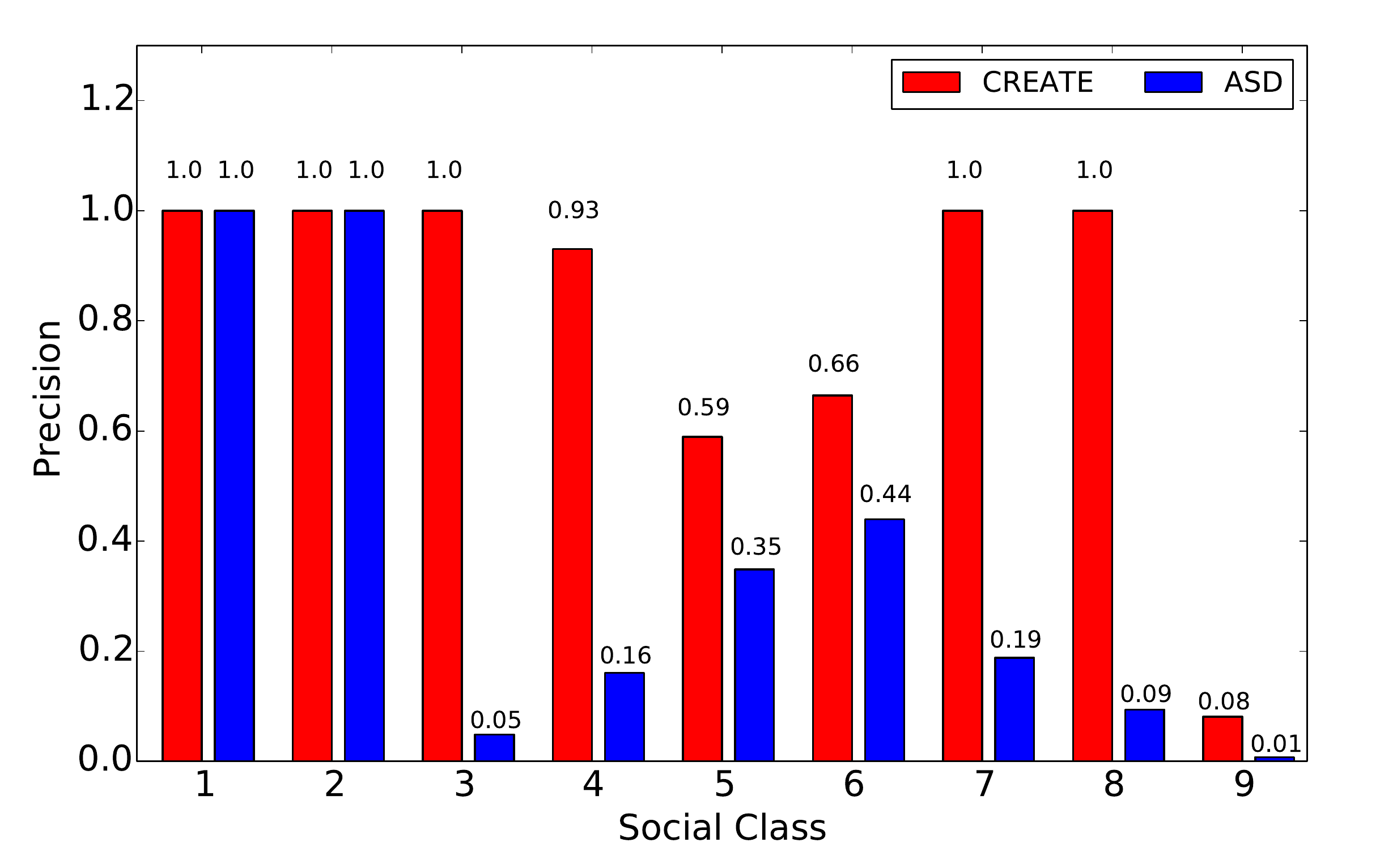}
    \end{minipage}
}
\subfigure[Recall]{ \label{eg_fig_comp_level_2}
    \begin{minipage}[l]{1.0\columnwidth}
      \centering
      \includegraphics[width=1.0\textwidth]{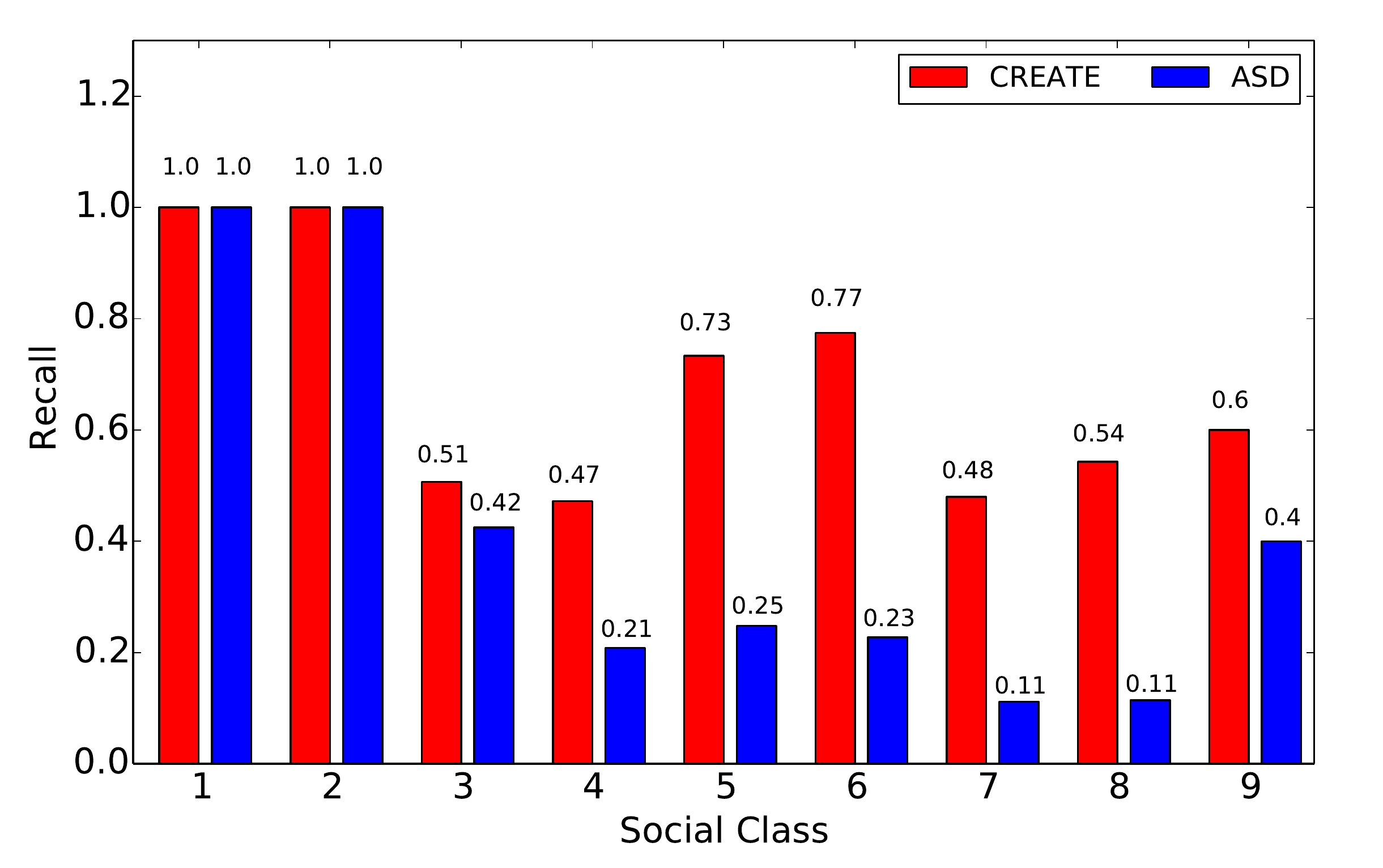}
    \end{minipage}
}
\caption{Precision and Recall achieved by {\our} and {\agony} at each social class of the organizational chart.}\label{eg_fig9_comp_level}
\end{figure*}

\begin{figure*}[t]
\centering
\subfigure[Accuracy]{\label{eg_fig_comp1_1}
    \begin{minipage}[l]{0.48\columnwidth}
      \centering
      \includegraphics[width=1.0\textwidth]{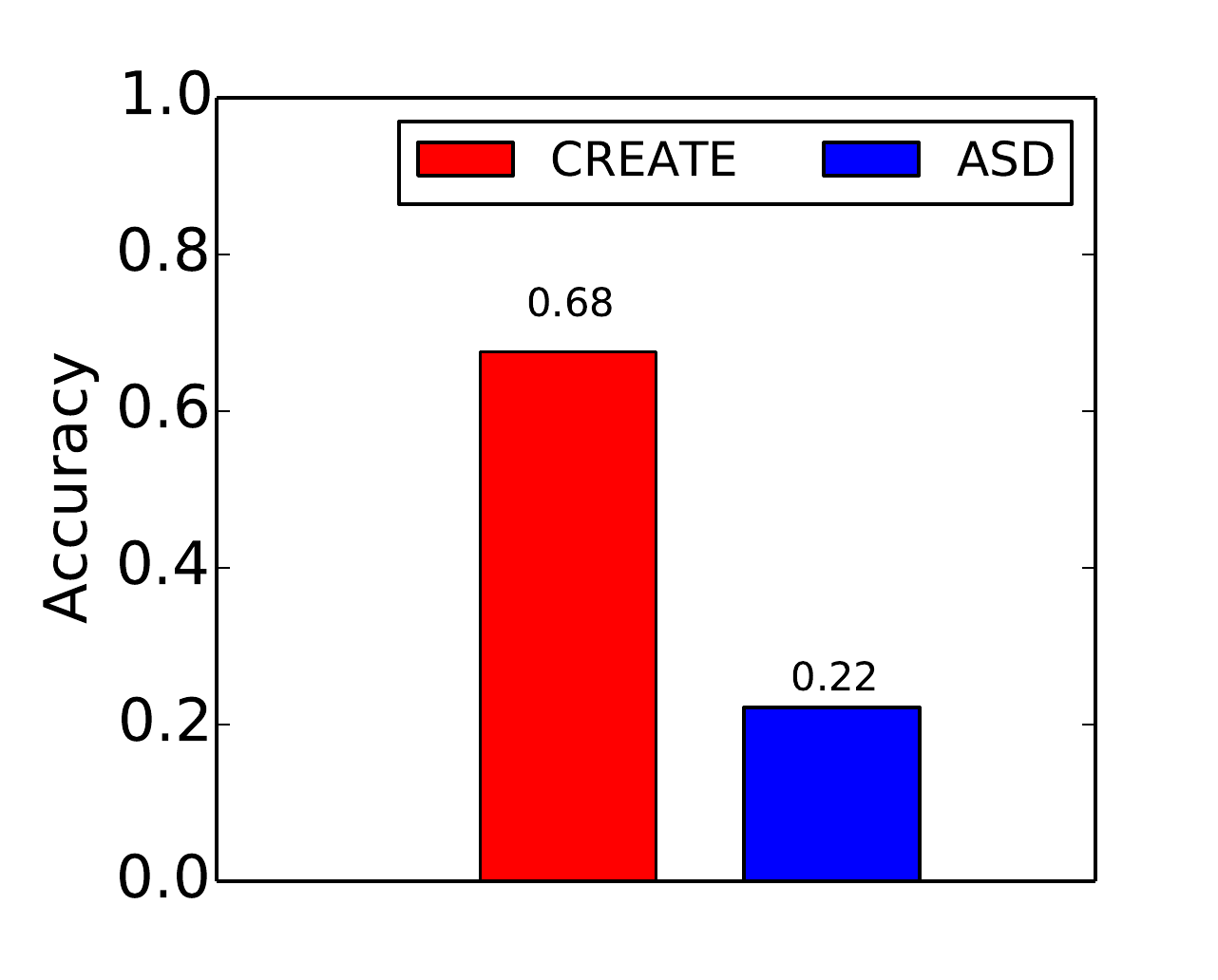}
    \end{minipage}
}
\subfigure[MAE]{ \label{eg_fig_comp1_2}
    \begin{minipage}[l]{0.48\columnwidth}
      \centering
      \includegraphics[width=1.0\textwidth]{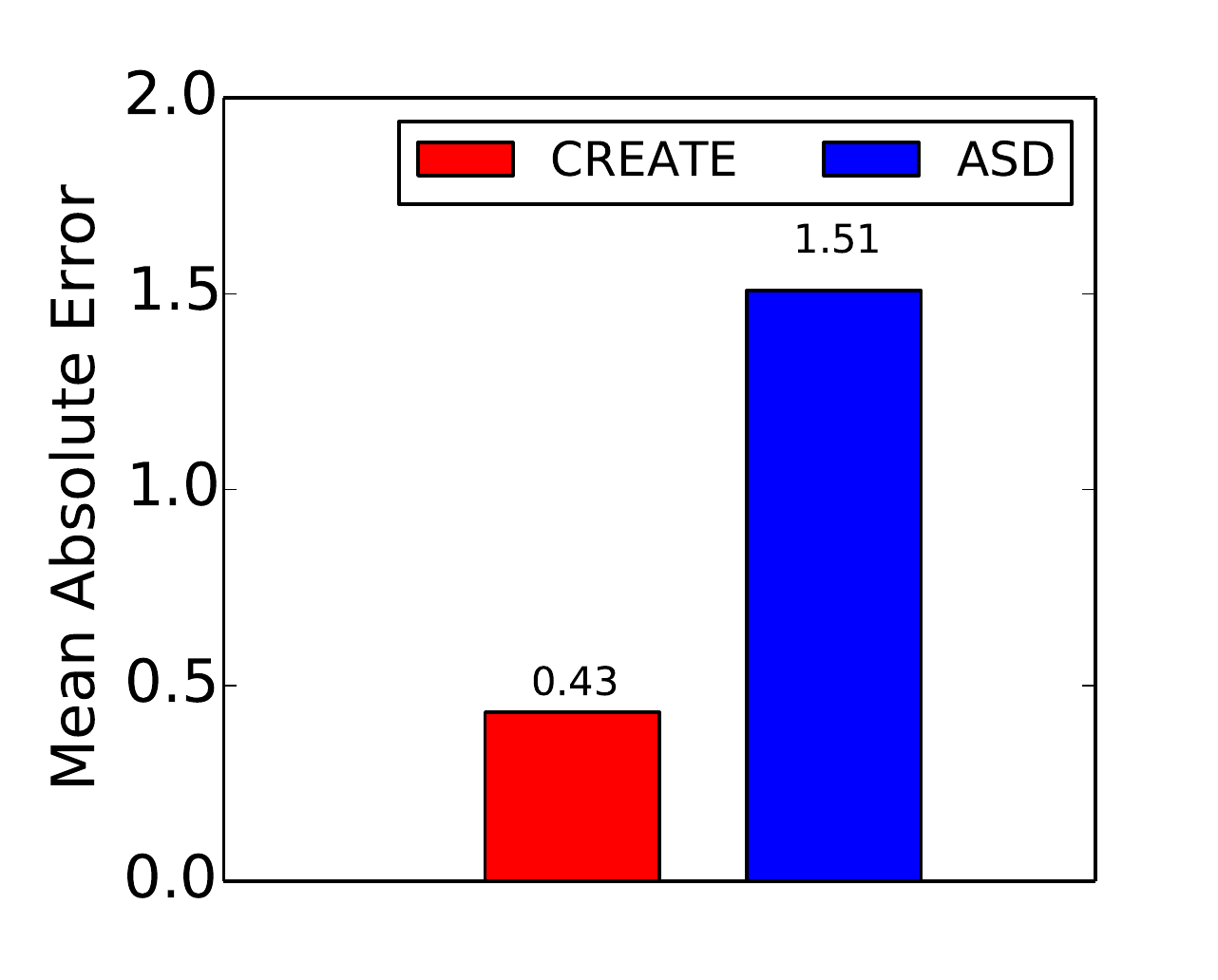}
    \end{minipage}
}
\subfigure[MSE]{ \label{eg_fig_comp1_3}
    \begin{minipage}[l]{0.48\columnwidth}
      \centering
      \includegraphics[width=1.0\textwidth]{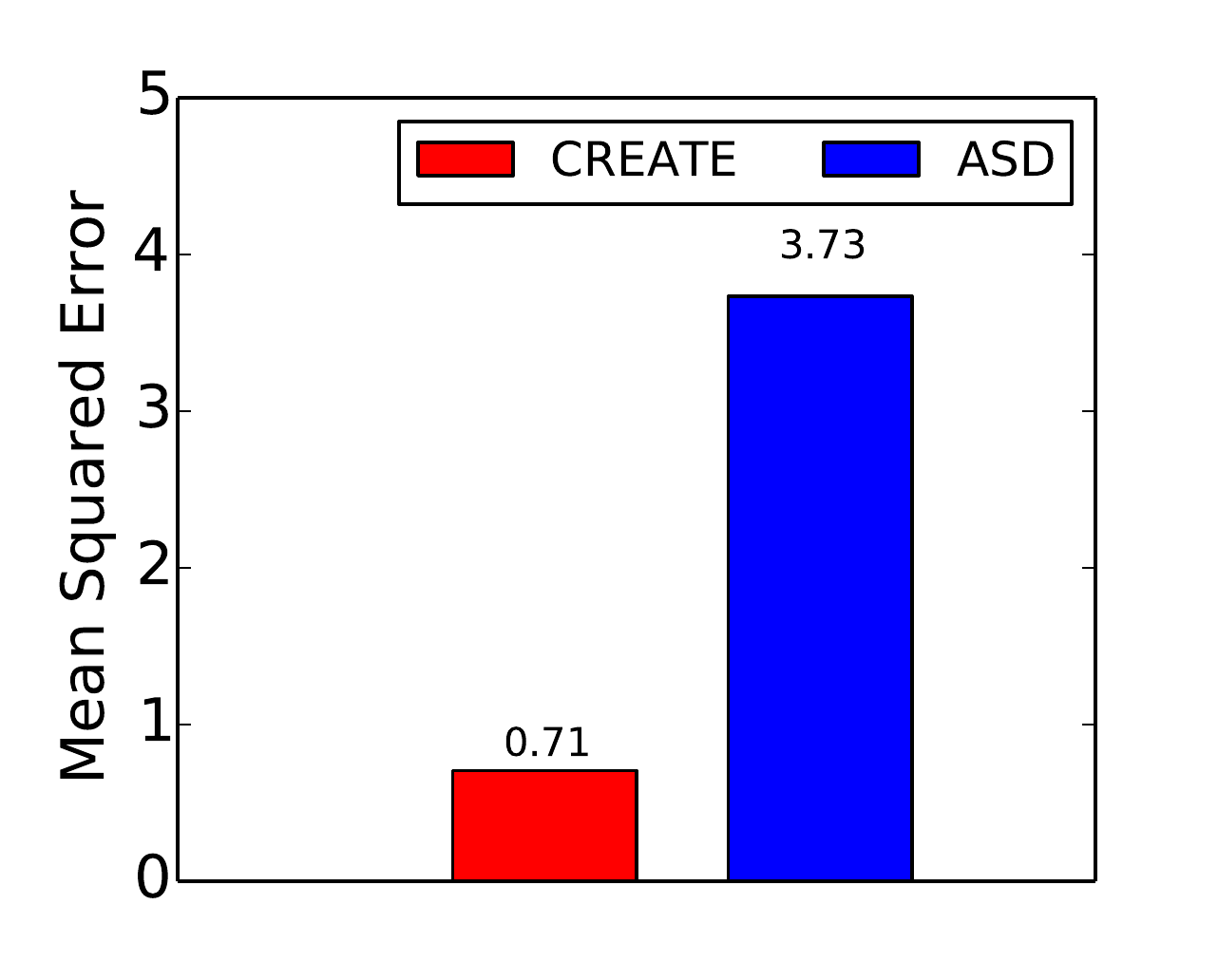}
    \end{minipage}
}
\subfigure[$R^2$]{ \label{eg_fig_comp1_4}
    \begin{minipage}[l]{0.48\columnwidth}
      \centering
      \includegraphics[width=1.0\textwidth]{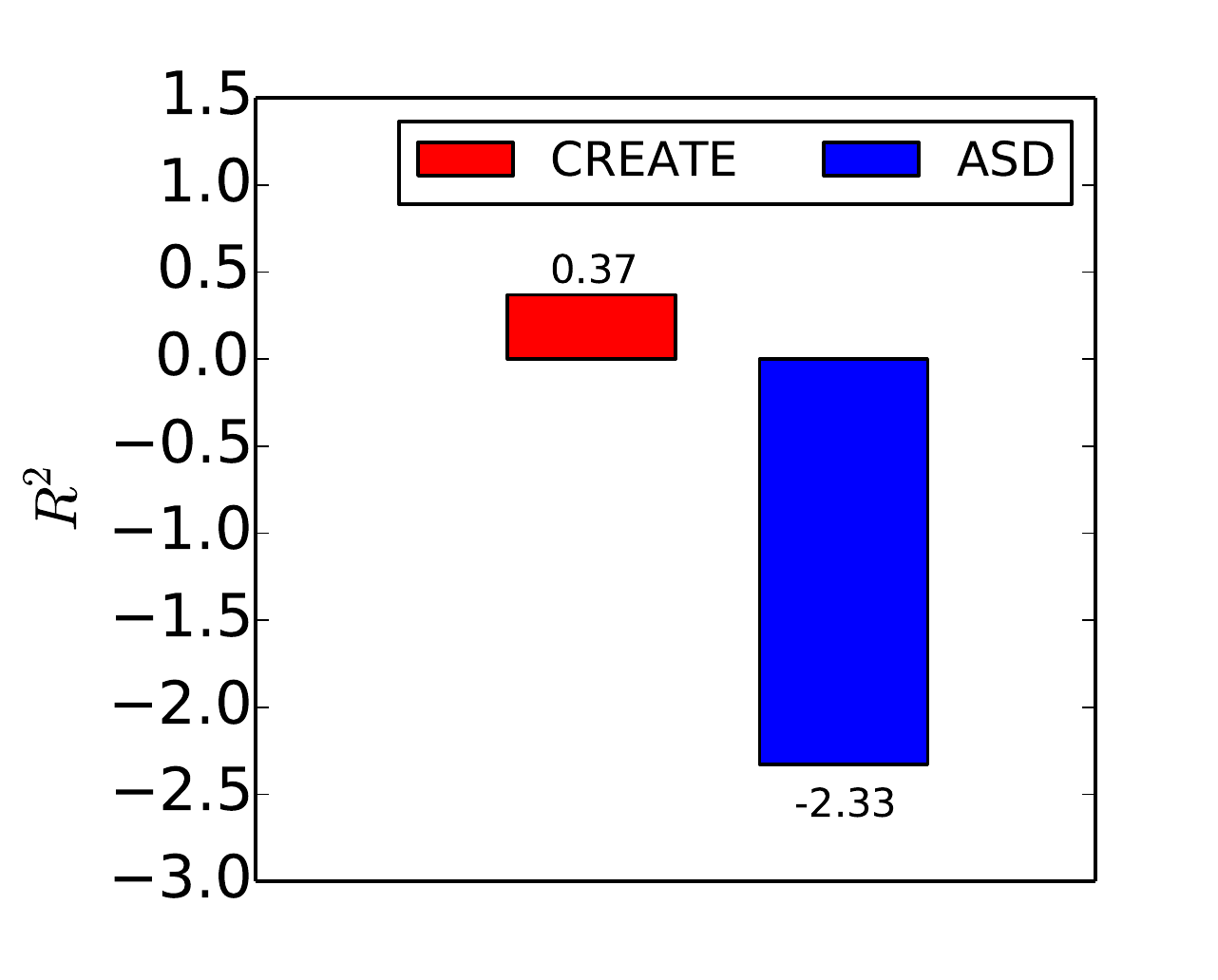}
    \end{minipage}
}
\caption{Performance comparison of {\our} and {\agony} evaluated by different metrics.}\label{eg_fig_comp1}
\end{figure*}

Social stratification results of {\our} and {\agony} are given in Figures~\ref{eg_fig9_comp_level}-\ref{eg_fig_comp1}, where Figure~\ref{eg_fig9_comp_level} shows the results achieved by {\our} and {\agony} at each social class (evaluated by precision and recall respectively) and Figure~\ref{eg_fig_comp1} shows their overall performance (evaluated by Accuracy, MAE, MSE and $R^2$).

From the microscopic perspective, we observes that {\our} performs better than {\agony} consistently at all social classes. {\our} achieves both $1.0$ precision and $1.0$ recall at social classes $1, 2$, i.e., {\our} identifies the top $2$ management levels of the company correctly. The performance of {\our} at other social classes is also very promising. For instance, the precision scores achieved by {\our} at social classes $3, 4, 7, 8$ (besides $1, 2$) are either $1.0$ or close to $1.0$ and the recall scores of {\our} at social classes $5, 6$ are also very high, which all outperform those of {\agony} with significant advantages.

From a macroscopic perspective, the performance of {\our} in stratifying the whole user set in ESN is very excellent and much better than that of {\agony}. The Accuracy, MAE, MSE and $R^2$ scores achieved by {\our}  are $0.68$, $0.43$, $0.71$ and $0.37$ respectively, which all outperforms those achieved by {\agony}. For example, the accuracy achieved {\our} is almost the triple of that obtained by {\agony}, while the MAE and MSE obtained by {\our} are merely the $28\%$ and $19\%$ of those achieved by {\agony}. In addition, {\agony} gets negative $R^2$ scores in identifying social classes of users in ESNs, which denotes that the identified users' social classes are massively disordered and have no linear correlation with the social class ground truth at all.

\subsection{Organizational Chart Inference Results}

\begin{table}[t]

\caption{Performance comparison of different organizational chart inference methods.}
\label{tab:result1}
\centering
{
\begin{tabular}{lcc}
\toprule
\multicolumn{1}{l}{ \multirow{2}{*}{Method} }&\multicolumn{2}{c}{Metrics}\\
\cmidrule{2-3}
&AUC	&Precision@100\\
\midrule
\midrule
{\our (K = 10)}	&0.856	&0.830	\\
{\our (K = 15)}	&\textbf{0.869}	&\textbf{0.870}	\\
{\our (K = 20)}	&\textbf{0.869}	&\textbf{0.870}	\\
{\oursl}	&0.719	&0.820	\\
{\oursm (K = 10)}	&0.610	&0.720	\\
{\oursm (K = 15)}	&0.630	&0.790	\\
{\oursm (K = 20)}	&0.630	&0.790	\\
{\ours}	&0.627	&0.740	\\
S-CN	&0.636	&0.440	\\
S-JC	&0.636	&0.260	\\
S-AA	&0.528	&0.070	\\

\bottomrule

\end{tabular}
}
\end{table}

\begin{figure*}[t]
\centering
\subfigure[Recall]{ \label{eg_fig_comp2_2}
    \begin{minipage}[l]{0.57\columnwidth}
      \centering
      \includegraphics[width=1.0\textwidth]{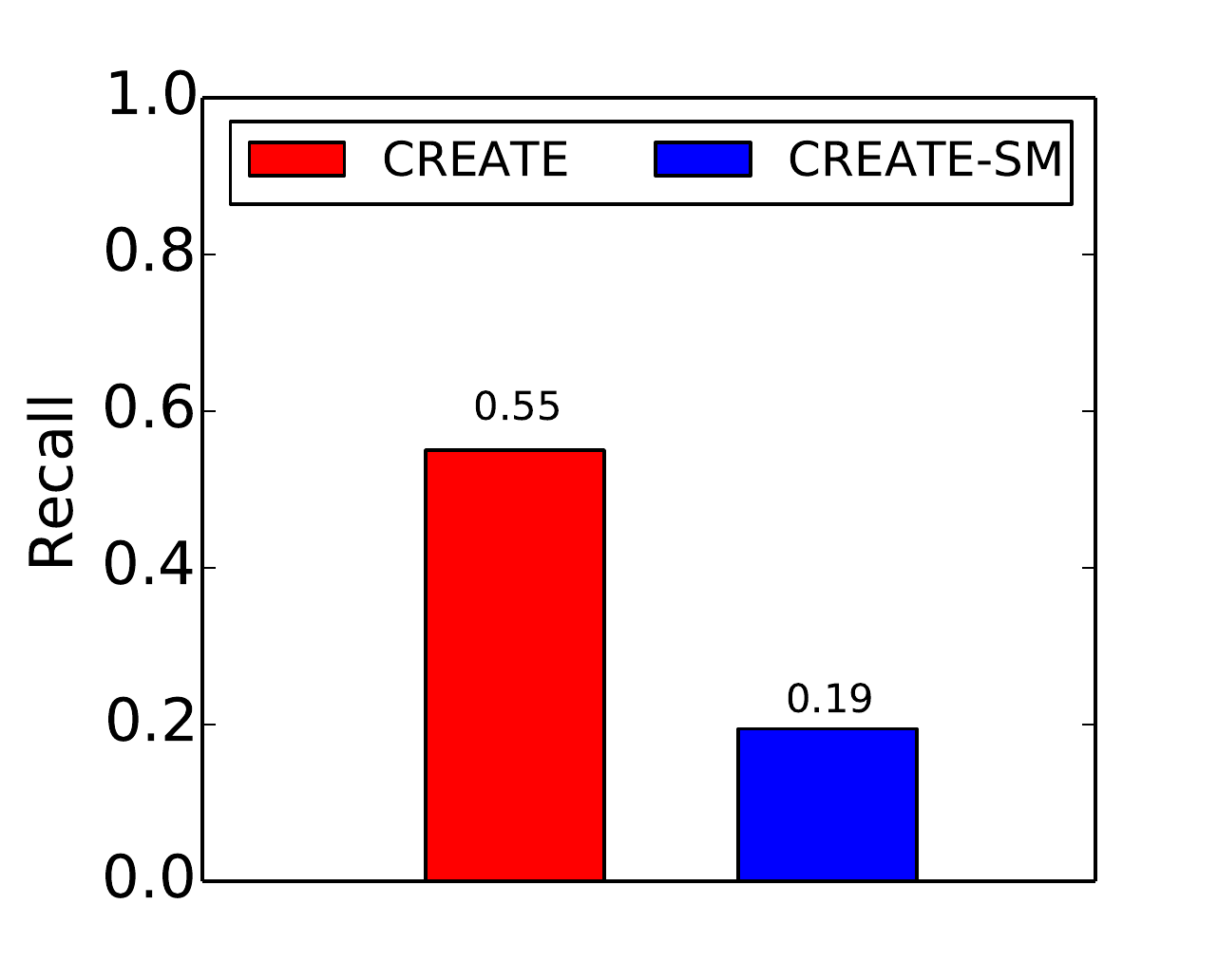}
    \end{minipage}
}
\subfigure[Precision]{ \label{eg_fig_comp2_3}
    \begin{minipage}[l]{0.57\columnwidth}
      \centering
      \includegraphics[width=1.0\textwidth]{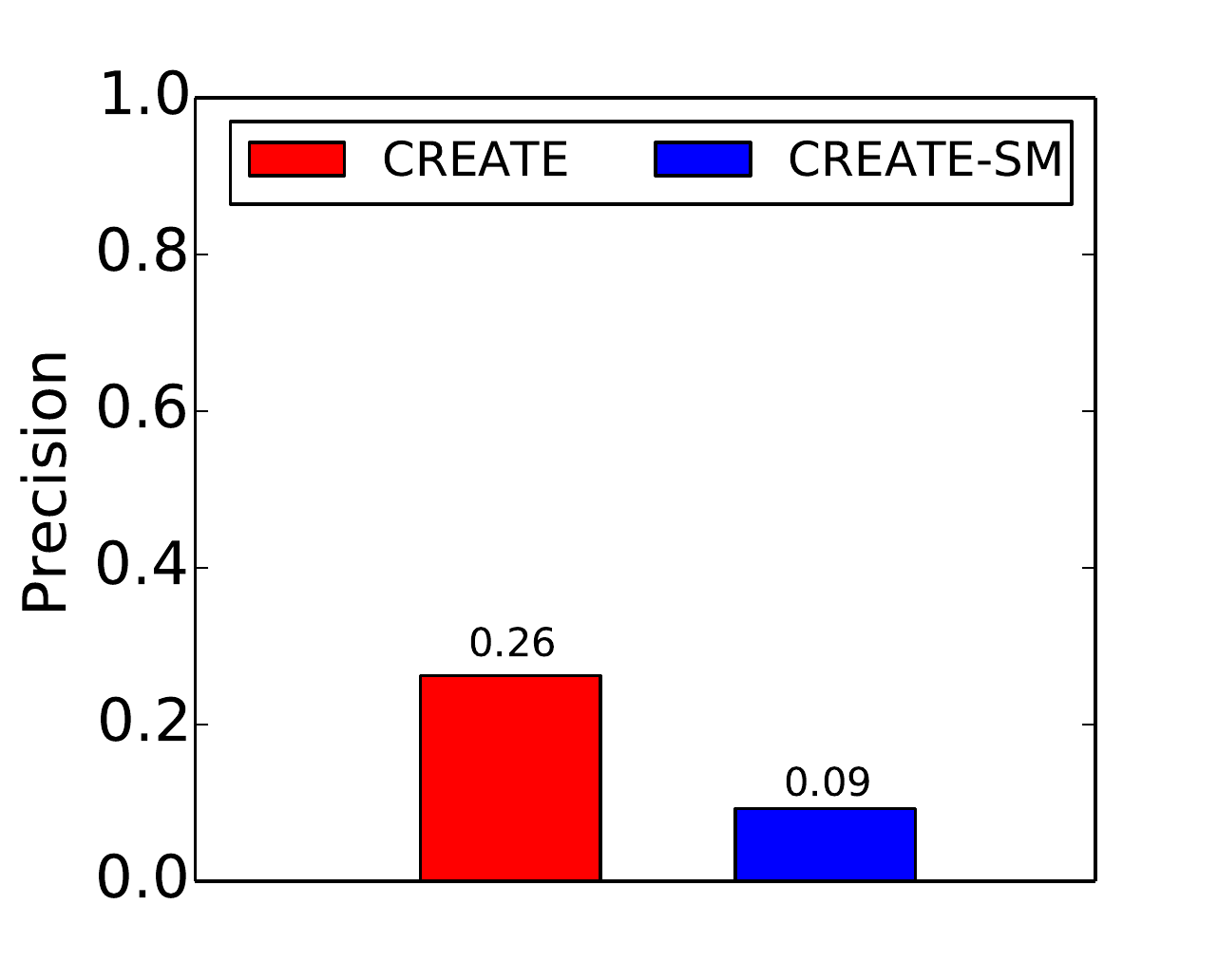}
    \end{minipage}
}
\subfigure[F1]{ \label{eg_fig_comp2_4}
    \begin{minipage}[l]{0.57\columnwidth}
      \centering
      \includegraphics[width=1.0\textwidth]{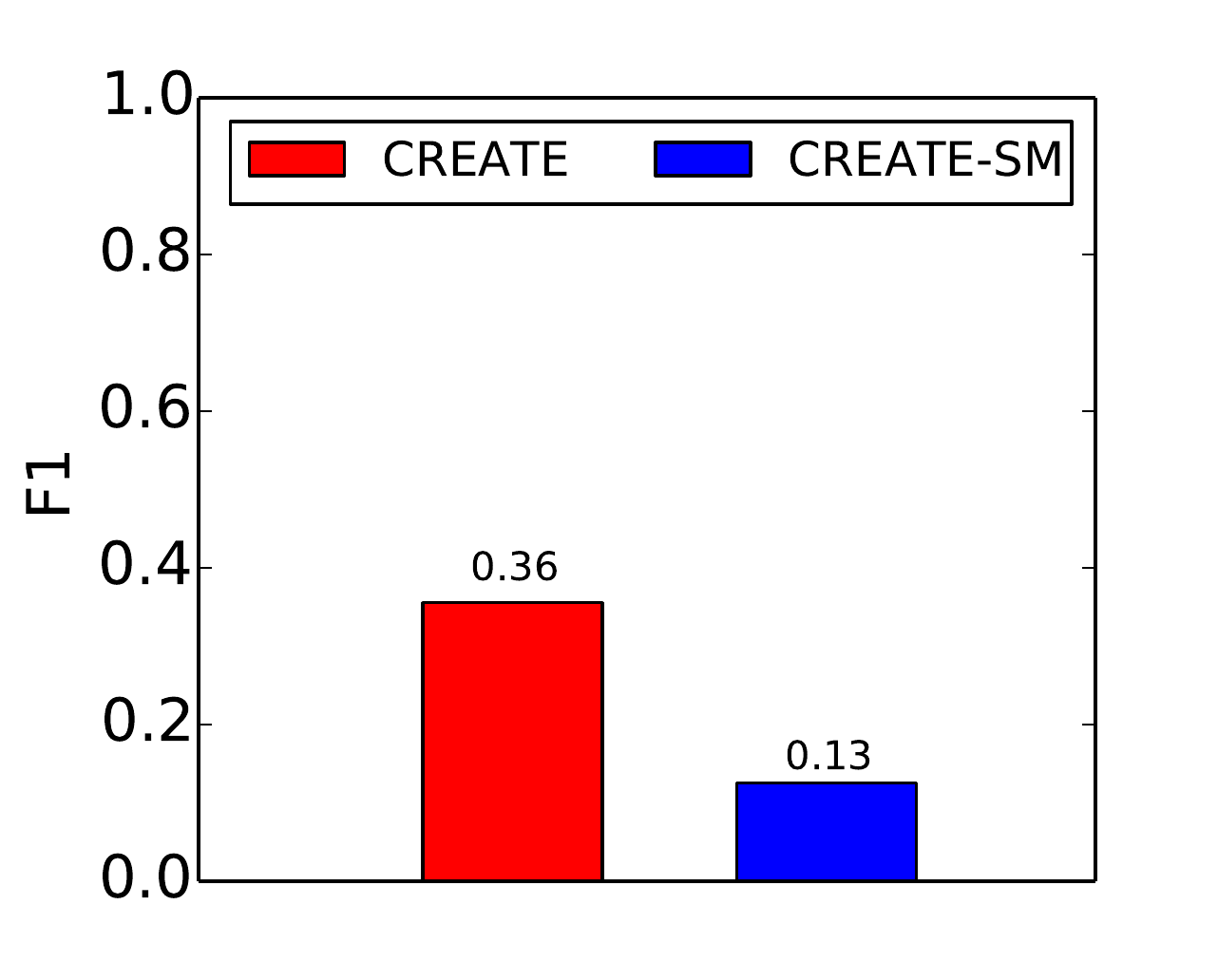}
    \end{minipage}
}
\caption{Performance comparison of {\our} and {\oursm} evaluated by different metrics (K = 15).}\label{eg_fig_comp2}
\end{figure*}

{\our} has proved its excellent effectiveness in stratifying users in ESNs, based on which, we further study its performance in inferring the potential supervision links between pairs of consecutive social classes, whose results are evaluated by AUC, Precision@100 in Table~\ref{tab:result1} as well as by Recall, Precision and F1 in Figure~\ref{eg_fig_comp2}.


In Table~\ref{tab:result1}, we compare {\our} (of different management thresholds $K$) with all the other baseline methods, where {\our} (with parameter $K = 15$ and $20$) performs the best. Compared with {\oursl} (or {\ours}), {\our} (or {\oursm}) which has the matching step can identify supervision links more effectively. For instance, {\our} (with $K=15$) outperforms {\oursl} by over $20\%$ in AUC and $6\%$ in Precision@100, and {\oursm} (with parameter $K=15$) outperforms {\ours} with remarkable advantages. It demonstrates that matching step can effectively prune non-existing supervision links and preserve the micro-level width requirement (i.e., the \textit{K-to-one} constraint).

Compared with {\oursm} and {\ours}, {\our} which infers the potential supervision links based on heterogeneous information in ESNs instead of merely regarding the social links as supervision link candidates achieves much better results. For example, in Table~\ref{tab:result1}, the AUC of {\our} is $38\%$ higher than that of {\oursm} and {\ours}, while the Precision@100 of {\our} is also roughly $10\%$ higher as well. In addition, in Figures~\ref{eg_fig_comp2}, the Recall, Precision and F1 obtained by {\our} is almost triple of those achieved by {\oursm}. It confirms the argument that heterogeneous information in ESNs can capture the relationships among colleagues (especially between managers and subordinates).

In addition, we also compare {\our} with traditional unsupervised link prediction methods, including CN, JC and AA, and the advantages of {\our} are very obvious according to Table~\ref{tab:result1}: {\our} can outperform all these unsupervised link prediction methods with significant advantages.

\subsection{Management Threshold Sensitivity Analysis}

\begin{figure*}[t]
\centering
\subfigure[Recall]{ \label{eg_fig_k_2}
    \begin{minipage}[l]{0.57\columnwidth}
      \centering
      \includegraphics[width=1.0\textwidth]{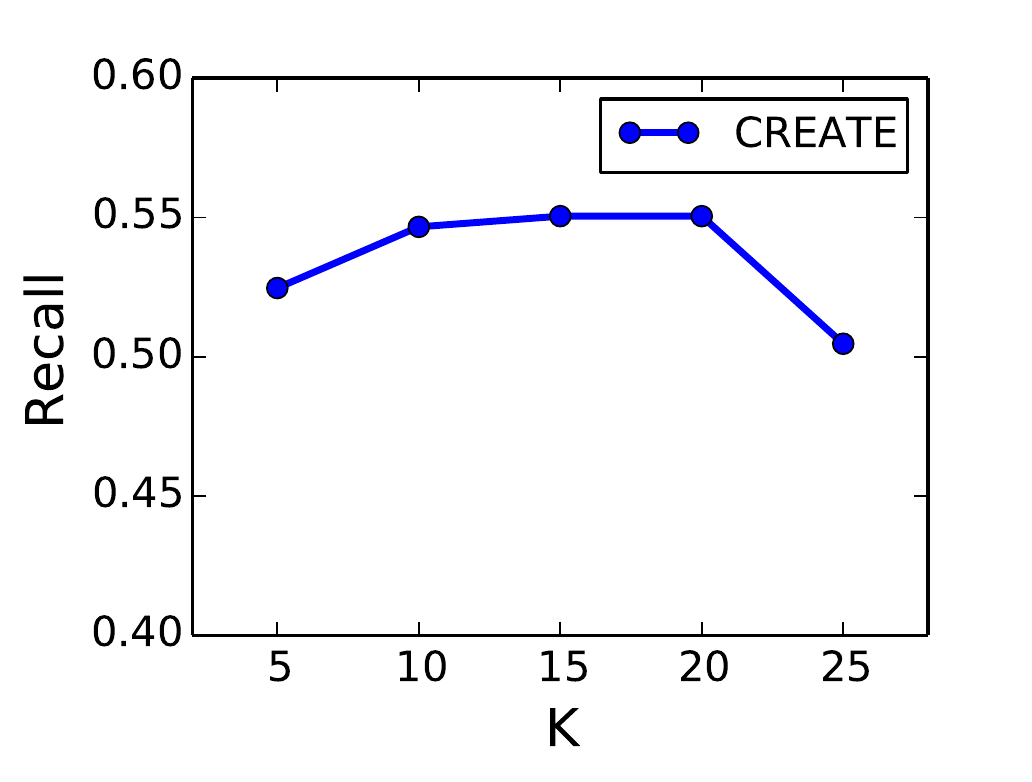}
    \end{minipage}
}
\subfigure[Precision]{ \label{eg_fig_k_3}
    \begin{minipage}[l]{0.57\columnwidth}
      \centering
      \includegraphics[width=1.0\textwidth]{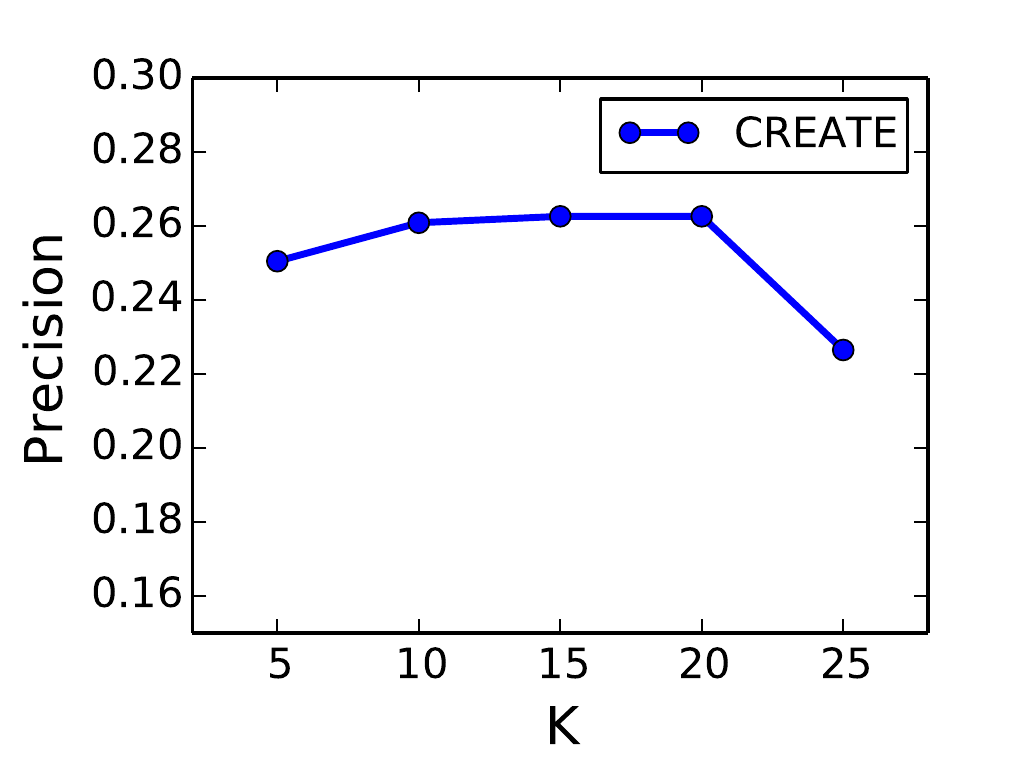}
    \end{minipage}
}
\subfigure[F1]{ \label{eg_fig_k_4}
    \begin{minipage}[l]{0.57\columnwidth}
      \centering
      \includegraphics[width=1.0\textwidth]{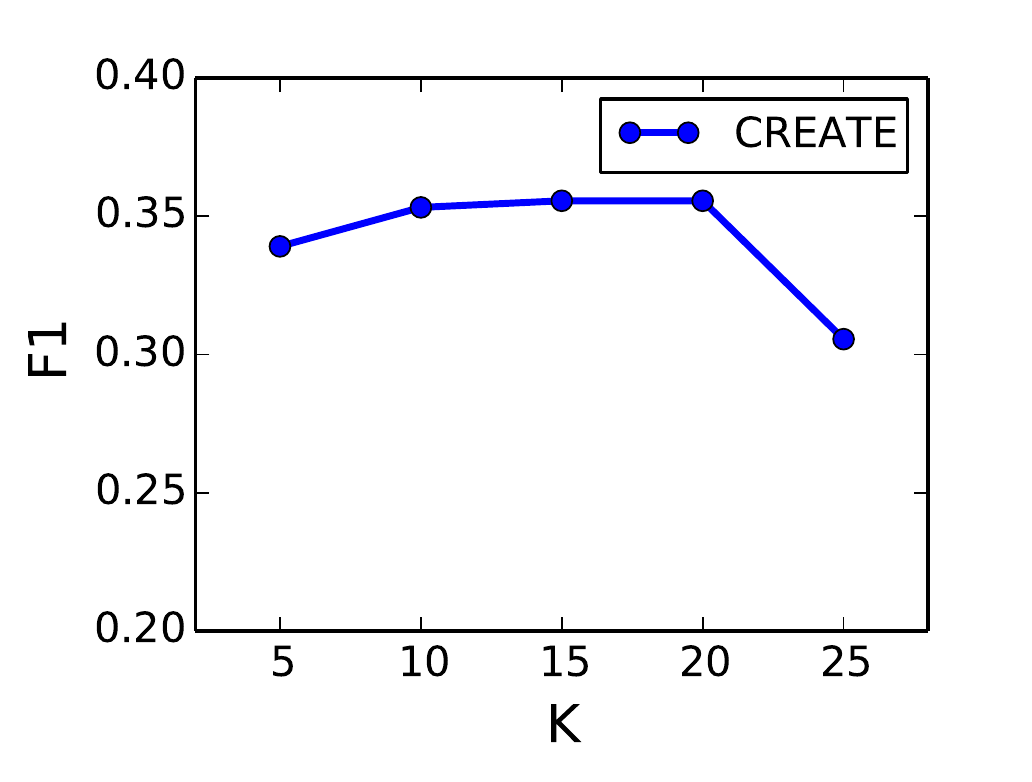}
    \end{minipage}
}
\caption{Sensitivity analysis of parameter K.}\label{eg_fig_k}
\end{figure*}

In social class matching, the management threshold parameter $K$ plays a key role in constraining the number of supervision links connected to each managers. The sensitivity of parameter $K$ will be analyzed in this section, where the results achieved by {\our} (with different $K$s) evaluated by different metrics are available in Figure~\ref{eg_fig_k}. Small management threshold $K$ (e.g., $5$) limits each manager's subordinate number to $5$ and will preserve the supervision links with extremely high likelihood only but may miss many promising ones. However, as threshold $K$ increases, more links with high likelihood will be preserved and the metric scores increase consistently. Meanwhile, when the threshold $K$ goes to $25$, the performance of {\our} degrades dramatically. The possible reason can be that, with larger threshold, each manager can have too many supervision links, which may exceed the subordinates they have in the real-world. 


\section{Related Work} \label{sec:relatedwork}

Enterprise social networks are important sources for employees in companies to get reliable information. Ehrlich et al. \cite{ELG07} propose to search for experts in enterprise with both text and social network analysis techniques. They propose to examine the users' dynamic profile information and get the social distance to the expert before deciding how to initiate the contact. Enterprise social networks can lead to lots of benefits to companies and the motivations of enterprise social network adoption in companies are studied in details in \cite{DMGDBM08}. Users in enterprise social networks will connect and learn from each other through personal and professional sharing. People sensemaking and relation building on an enterprise social network site is studied in by DiMicco et al. \cite{DGMDB09}. In addition, social connections among users in enterprise social networks usually have multiple facets. Wu et al. \cite{WDM10} propose to study the study the multiplexity of social connections among users in enterprise social networks, which include both professional and personal closeness. 


From social networks, some works have been done to infer the hierarchies of individuals \cite{CMN07, MB09, GSLMI11, JWPH14}. A measure, agony, is proposed in \cite{GSLMI11}, by minimizing which the authors propose a hierarchy detection method. A random graph model and markov chain monte carlo sampling is proposed by Clauset et al. in \cite{CMN07}, which can address the problem of structural inference of hierarchies in networks. Maiya et al. propose to identify the hierarchies in social networks to achieve the maximum likelihood in \cite{MB09}. All the above three papers focus on dividing individuals into different hierarchies only. A offline hierarchical ties inference method has been proposed by Jaber et al. in \cite{JWPH14} to discover offline links among people based on a time-based model. However, none of these papers can recover the whole organizational chart.

Cross-social-network studies has become a hot research topic in recent years. Kong et al. \cite{KZY13} are the first to propose the concepts of ``anchor links'', ``anchor users'', ``'aligned networks'' etc. A novel network anchoring method is proposed in \cite{KZY13} to address the network alignment problem. Cross-network heterogeneous link prediction problems are studied by Zhang et al. \cite{ZKY13, ZKY14, ZYZ14, ZY15-3} by transferring links across partially aligned networks. Besides link prediction problems, Jin et al. proposes to partition multiple large-scale social networks simultaneously in \cite{JZYYL14} and Zhang et al. study the community detection problem across partially aligned networks in \cite{ZY15, ZY15-2}. Zhan et al. analyze the information diffusion process across aligned networks \cite{ZZWYX15}.


\section{Conclusion}\label{sec:conclusion}

In this paper, we have studied the organizational chart inference ({\problem}) problem based on the heterogeneous online ESNs. To address the {\problem} problem, a new chart inference framework {\our} has been proposed in Section~\ref{sec:method}. {\our} consists of $3$ steps: (1) regulated social stratification, (2) supervision link inference with social meta paths aggregation, and (3) regulated social class matching. Experiments on real-world ESN and organizational chart dataset have demonstrated the effectiveness of {\our}.

\label{sec:ack}
\section{Acknowledgement}
This work is supported in part by NSF through grants CNS-1115234, Google Research Award, the Pinnacle Lab at Singapore Management University, and Huawei grants.


\begin{thebibliography}{10}

\bibitem{boss}
Be the boss, not a friend.
\newblock
  \url{http://www.pereg.com/manager/FCKeditor/editor/filemanager/connectors/aspx/pereg.com/userfiles/file/Be%20the%20boss.pdf}.
\newblock [Online; accessed 24-December-2014].

\bibitem{org}
Organisation-organizational structure-organisational chart.
\newblock
  \url{http://kalyan-city.blogspot.com/2010/06/organisation-organizational-structure.html
  }.
\newblock [Online; accessed 29-January-2015].

\bibitem{life}
Rich, poor, and middle class life.
\newblock
  \url{http://www.historydoctor.net/Advanced%20Placement%20European%20History/Notes/rich_poor_and_middle_class_life.htm}.
\newblock [Online; accessed 24-December-2014].

\bibitem{fasion}
The social foundations of fashion change in the late 19th century.
\newblock \url{http://www.marquise.de/en/1800/ch1800soz.shtml}.
\newblock [Online; accessed 24-December-2014].

\bibitem{AA01}
L.~Adamic and E.~Adar.
\newblock Friends and neighbors on the web.
\newblock {\em Social Networks}, 2001.

\bibitem{A09}
M.~Aoki.
\newblock Horizontal vs. vertical information structure of the firm.
\newblock {\em The American Economic Review}, 2009.

\bibitem{CMN07}
A.~Clauset, C.~Moore, and M.~Newman.
\newblock Structural inference of hierarchies in networks.
\newblock In {\em Statistical Network Analysis: Models, Issues, and New
  Directions}. 2007.

\bibitem{D97}
R.~Diestel.
\newblock {\em Graph Theory}.
\newblock Springer, 1997.

\bibitem{DGMDB09}
J.~DiMicco, W.~Geyer, D.~Millen, C.~Dugan, and B.~Brownholtz.
\newblock People sensemaking and relationship building on an enterprise social
  network site.
\newblock In {\em HICSS}, 2009.

\bibitem{DMGDBM08}
J.~DiMicco, D.~Millen, W.~Geyer, C.~Dugan, B.~Brownholtz, and M.~Muller.
\newblock Motivations for social networking at work.
\newblock In {\em CSCW}, 2008.

\bibitem{ELG07}
K.~Ehrlich, C.~Lin, and V.~Griffiths-Fisher.
\newblock Searching for experts in the enterprise: Combining text and social
  network analysis.
\newblock In {\em GROUP}, 2007.

\bibitem{EFKE12}
A.~Elishar, M.~Fire, D.~Kagan, and Y.~Elovici.
\newblock Organizational intrusion: Organization mining using socialbots.
\newblock In {\em SocialInformatics}, 2012.

\bibitem{G01}
J.~Grant.
\newblock Class, definition of.
\newblock In R.J.~Barry Jones, editor, {\em Routledge Encyclopedia of
  International Political Economy}. 2001.

\bibitem{GSHB14}
H.~Gui, Y.~Sun, J.~Han, and G.~Brova.
\newblock Modeling topic diffusion in multi-relational bibliographic
  information networks.
\newblock In {\em CIKM}, 2014.

\bibitem{GSLMI11}
M.~Gupte, P.~Shankar, J.~Li, S.~Muthukrishnan, and L.~Iftode.
\newblock Finding hierarchy in directed online social networks.
\newblock In {\em WWW}, 2011.

\bibitem{HZ11}
M.~Hasan and M.~J. Zaki.
\newblock A survey of link prediction in social networks.
\newblock In Charu~C. Aggarwal, editor, {\em Social Network Data Analytics}.
  2011.

\bibitem{H12}
S.~Hill.
\newblock Elite and upper-class families.
\newblock In {\em Families: A Social Class Perspective}. 2012.

\bibitem{HK06}
R.~Hyndman and A.~Koehler.
\newblock {Another look at measures of forecast accuracy}.
\newblock {\em IJF}, 2006.

\bibitem{JWPH14}
M.~Jaber, P.~Wood, P.~Papapetrou, and S.~Helmer.
\newblock Inferring offline hierarchical ties from online social networks.
\newblock In {\em WWW Companion}, 2014.

\bibitem{JZYYL14}
S.~Jin, J.~Zhang, P.~Yu, S.~Yang, and A.~Li.
\newblock Synergistic partitioning in multiple large scale social networks.
\newblock In {\em IEEE BigData}, 2014.

\bibitem{KZY13}
X.~Kong, J.~Zhang, and P.~Yu.
\newblock Inferring anchor links across heterogeneous social networks.
\newblock In {\em CIKM}, 2013.

\bibitem{K02}
V.~Krebs.
\newblock Mapping networks of terrorist cells.
\newblock {\em Connections}, 2002.

\bibitem{K95}
A.~Kroch.
\newblock Dialect and style in the speech of upper class philadelphia.
\newblock In {\em Towards a Social Science of Language: Papers in honor of
  William Labov}. 1995.

\bibitem{MB09}
A.~Maiya and T.~Berger-Wolf.
\newblock Inferring the maximum likelihood hierarchy in social networks.
\newblock In {\em CSE}, 2009.

\bibitem{M88}
R.~Merton.
\newblock The matthew effect in science, ii: Cumulative advantage and the
  symbolism of intellectual property.
\newblock {\em ISIS}, 1988.

\bibitem{NBZ06}
N.~Nagappan, T.~Ball, and A.~Zeller.
\newblock Mining metrics to predict component failures.
\newblock In {\em ICSE}, 2006.

\bibitem{RPD88}
J.~Rawlings, S.~Pantula, and D.~Dickey.
\newblock {\em Applied Regression Analysis}.
\newblock 1988.

\bibitem{SBGAH11}
Y.~Sun, R.~Barber, M.~Gupta, C.~Aggarwal, and J.~Han.
\newblock Co-author relationship prediction in heterogeneous bibliographic
  networks.
\newblock In {\em ASONAM}, 2011.

\bibitem{SHYYW11}
Y.~Sun, J.~Han, X.~Yan, P.~Yu, and T.~Wu.
\newblock Pathsim: Meta path-based top-k similarity search in heterogeneous
  information networks.
\newblock In {\em VLDB}, 2011.

\bibitem{US07}
US~Army Training and Doctrine Command.
\newblock {\em Terrorist Organizational Models}.
\newblock 2007.

\bibitem{WBSS04}
Z.~Wang, A.~Bovik, H.~Sheikh, and E.~Simoncelli.
\newblock Image quality assessment: From error visibility to structural
  similarity.
\newblock {\em IEEE IP}, 2004.

\bibitem{W13}
L.~Warwick-Booth.
\newblock {\em Social Inequality: A Student's Guide}.
\newblock SAGE Publications Ltd, 2013.

\bibitem{WDM10}
A.~Wu, J.~DiMicco, and D.~Millen.
\newblock Detecting professional versus personal closeness using an enterprise
  social network site.
\newblock In {\em CHI}, 2010.

\bibitem{ZZWYX15}
Q.~Zhan, S.~Wang J.~Zhang, P.~Yu, and J.~Xie.
\newblock Influence maximization across partially aligned heterogenous social
  networks.
\newblock In {\em PAKDD}, 2015.

\bibitem{ZKY13}
J.~Zhang, X.~Kong, and P.~Yu.
\newblock Predicting social links for new users across aligned heterogeneous
  social networks.
\newblock In {\em ICDM}, 2013.

\bibitem{ZKY14}
J.~Zhang, X.~Kong, and P.~Yu.
\newblock Transferring heterogeneous links across location-based social
  networks.
\newblock In {\em WSDM}, 2014.

\bibitem{ZY15}
J.~Zhang and P.~Yu.
\newblock Community detection for emerging networks.
\newblock In {\em SDM}, 2015.

\bibitem{ZY15-3}
J.~Zhang and P.~Yu.
\newblock Integrated anchor and social link predictions across partially
  aligned social networks.
\newblock In {\em IJCAI}, 2015.

\bibitem{ZY15-2}
J.~Zhang and P.~Yu.
\newblock Mcd: Mutual clustering across multiple heterogeneous networks.
\newblock In {\em IEEE BigData Congress}, 2015.

\bibitem{ZYZ14}
J.~Zhang, P.~Yu, and Z.~Zhou.
\newblock Meta-path based multi-network collective link prediction.
\newblock In {\em KDD}, 2014.

\end{thebibliography}
\end{document}